\documentclass[fleqn,usenatbib]{mnras}

\usepackage{newtxtext,newtxmath}

\usepackage[T1]{fontenc}

\DeclareRobustCommand{\VAN}[3]{#2}
\let\VANthebibliography\thebibliography
\def\thebibliography{\DeclareRobustCommand{\VAN}[3]{##3}\VANthebibliography}

\usepackage{graphicx}	
\usepackage{amsmath}	

\usepackage{amssymb}	

\usepackage{threeparttable, tablefootnote}

\usepackage{lscape}

\title[X-ray emission of NGC 4051]{Investigation on the X-ray emission of NGC 4051 during its 2009 optical/UV-X-ray dissociation phase}

\author[M. Zhou et al.]{Minhua Zhou,$^{1}$\thanks{E-mail: zhoumh8@163.com (MZ)}
	Xinling Wu,$^{1}$
	Lei Xu$^{1}$ and
	Nannan Chen$^{2}$
	\\
	$^{1}$Jiangxi Province Key Laboratory of Applied Optical Technology, School of Physical Science and Intelligent Education, Shangrao Normal College, 401 Zhimin Road, Shangrao 334001, China\\
	$^{2}$School of Digital Technology Application Industry, Shangrao Normal College, 401 Zhimin Road, Shangrao 334001, People's Republic of China
}

\date{Accepted XXX. Received YYY; in original form ZZZ}

\pubyear{2026}

\begin{document}
\label{firstpage}
\pagerange{\pageref{firstpage}--\pageref{lastpage}}
\maketitle

\begin{abstract}

This study investigates the X-ray characteristics of jet-associated radio-quiet AGNs across distinct optical/UV to X-ray correlation phases. Quasi-simultaneous optical/UV/X-ray observations of NGC 4051 from May-June 2009, obtained through Swift and XMM-Newton, reveal a temporal dichotomy: a strong optical/UV to X-ray correlation dominates the initial observation phase (before May 27), followed by an optical/UV flare event concurrent with X-ray flux suppression in the latter period.
Our multi-method analysis of XMM-Newton data, incorporating short-term X-ray variability assessment, spectral decomposition, and RGS spectral analysis, identifies significant inter-phase X-ray emission disparities. During optical/UV flaring episodes, compared to the correlated phase, we observe: 
attenuated short-term X-ray variability amplitudes, 
enhanced soft X-ray absorption, 
suppressed intrinsic hard X-ray flux, and more prominent RGS emission-line features. 
Notably, these X-ray characteristics during optical/UV flaring intervals show no statistically significant deviations from pre-flare low-state X-ray emission patterns. 
These non-synchronous optical/UV-X-ray variations contradict predictions from both reprocessing models, starburst-driven emission scenarios, and the simplistic absorption models. 
While potential jet-related mechanisms remain ambiguous, our findings demonstrate strong consistency with predictions from the inhomogeneous accretion disk perturbation framework.

\end{abstract}




\section{Introduction}
\label{sec:intro}

Active galactic nuclei (AGN) are among the most luminous and energetic objects in the universe, powered by accretion onto supermassive black holes (e.g., Shakura \& Sunyaev 1973). 
Based on the radio observations, AGNs can be classified into radio-loud and radio-quiet AGNs with radio loudness \cite[the ratio of radio to optical luminosity, $R=f_{5\rm GHz}/f_{4400}$,][]{1989AJ.....98.1195K} $R \ge 10$ and $R<10$ in the former and latter, respectively. 
In the AGNs unified model \cite[][]{1993ARA&A..31..473A}, the optical/ultraviolet (UV) photons of radio-quiet AGNs (RQ-AGNs) originate from the thermal process of the accretion disk \cite[][]{1973A&A....24..337S}, and part of them are converted to the primary X-rays in an accretion ``corona'' by the inverse Compton process \cite[][]{1980A&A....86..121S, 1985A&A...143..374S}. 
For the presence of powerful relativistic jets in radio-loud AGNs (RL-AGNs), however, the radio emission of RL-AGNs is significantly different from that of RQ-AGNs \cite[][]{1995PASP..107..803U, 2019NatAs...3..387P}, and may even cause (directly or indirectly) the differences in the X-ray emission of these two radio dichotomy AGNs \cite[e.g.,][]{1987ApJ...313..596W, 2002MNRAS.332L..45B, 2011ApJ...726...20M, 2021RAA....21....4Z}. 

Variability, a hallmark of AGN emission across electromagnetic bands, offers unique insights into the physical processes governing these compact regions \cite[e.g., ][]{1997iagn.book.....P}. In particular, multi-wavelength flux variability correlations between the optical/UV and X-ray bands have emerged as a powerful diagnostic tool for constraining accretion disk-corona coupling mechanisms and the origins of X-ray emission \cite[e.g., ][]{2000ApJ...535...58K, 2001ApJ...561..162S}. 
Studies of optical/UV-X-ray correlations in AGN have revealed a complex picture. Some sources exhibit significant correlations with time lags consistent with the thermal Comptonization model \cite[e.g., ][]{2000ApJ...544..734N, 2019ApJ...870L..13A, 2023MNRAS.521.4109K}, or suggesting direct reprocessing of X-rays by the accretion disk \cite[e.g.,][]{2009MNRAS.394..427B, 2010MNRAS.403..605B, 2012MNRAS.422..902C, 2017ApJ...840...41E, 2023MNRAS.521.4109K}. Conversely, some AGN show weak or no detectable correlations \cite[e.g., ][]{2019ApJ...870...54M, 2022MNRAS.514.2974M, 2023ApJ...958...46B, 2023ApJ...958..195C}, challenging standard disk-corona paradigms and hinting at alternative scenarios, such as relativistic beaming emission \cite[e.g.,][]{2019ApJ...870...54M, 2023MNRAS.519..909Z} or spatially distinct variability drivers \cite[e.g.,][]{2018MNRAS.475.2306B} or inhomogeneous disc model \cite[e.g.,][]{2022MNRAS.512.5511S}. These discrepancies underscore the need for detailed case studies of individual AGN to unravel the diversity of physical processes at play.

\cite{2018MNRAS.475.2306B} found that the variability in the UV emission for IRAS 13224$-$3809 does not clearly correlate with that of X-rays over the available $\sim$40 d monitoring. The authors argued the the UV emitting outer disc does not correlate with the X-ray flux in our line of sight and/or that another process drives the majority of the UV variability. The former case may be due to changes in coronal geometry, absorption or scattering between the corona and the disc. 
This lack of correlation in IRAS 13224$-$3809 is also presented in 1H 0707$-$495 \cite[][]{2015MNRAS.453.3455R} with a much significant variation amplitude of the X-ray than that of the UV band. These results are consistent with the X-ray source being centrally compact and dominated by light bending close to the black hole. 
Based on the disk turbulence model \cite[][]{2018ApJ...855..117C}, the model in \cite{2020ApJ...892...63C} can well reproduce the observed UV to X-ray lags and the optical to UV lags simultaneously in four local Seyfert galaxies by assuming the corona heating is associated with turbulence in the inner accretion disk. 
If the X-ray and extreme-UV emissions has been obscured \cite[by disk winds, e.g.,][]{2019ApJ...877..119D} or altered by the turbulence disk \cite[with truncation radii comparable to the characteristic scales of UV emission zones,][]{2020ApJ...892...63C}, the framework proposed in \cite{2020ApJ...892...63C} could naturally account for the observed lack of correlation between X-ray and UV emission signatures. 






Considering the jet properties and variability characteristics, studying the X-ray emission in different correlation phases of jet-hosting radio-quiet AGN may further reveal the impact of jets on their multi-wavelength radiation and flux variations. 
NGC 4051 ($\alpha_{2000}=\rm 12^h03^m09.614^s$, $\delta_{2000}=\rm +44^d31^m52.80^s$, $z=0.002336$, $z=0.002336$), a nearby low-luminosity Seyfert galaxy and a documented radio-quiet AGN with jets \cite[e.g.,][]{2017MNRAS.465.1336J}, emerges as an ideal laboratory for investigating jet-related phenomena due to its extreme X-ray variability and complex spectral evolution observed across multiple epochs.
\cite{1990MNRAS.243..713D} did the first quasi-simultaneous infrared, optical, ultraviolet, and X-ray variability studies of NGC 4051, revealing that the amplitude of optical variability was significantly smaller than that of X-ray.
Combined with the historical data, \cite{2010MNRAS.403..605B} presented the significant correlation of an optical delay of $1.2^{+1.0}_{-0.3}\rm ~d$ behind the X-rays by using X-ray and optical light curves spanning more than 12 years (from 1996 to 2008), and confirmed that the amplitude of optical variability is much less than that of X-rays. It indicates the optical variations could be driven by X-ray reprocessing. 
However, \cite{2013MNRAS.429...75A} found no significant correlation between UV and X-ray emission with XMM-Newton and Swift data during May and June 2009. 
With AstroSat data, \cite{2024MNRAS.527.5668K} concluded that NGC 4051 has correlated multi-band variations with X-ray leading those in the FUV and NUV bands by $\sim7.4\rm ~ ks$ and $\sim24.2\rm~ ks$ during $5-9$ June 2016, respectively. 
Summarizing the above results, although NGC 4051 has complex multi-band variability, its optical/UV and X-ray continuum variations have a good correlation, except that during the XMM-Newton observation period in 2009. 
The unusual correlation between UV and X-ray emissions of NGC 4051 in 2009 remains to be discussed.

In this work, we investigate the X-ray emission of NGC 4051 during its different optical/UV to X-ray relation phase with XMM-Newton observations. 
The following Section presents our data analysis method. Our main results of NGC 4051 are shown in Section \ref{Sec:results}. 
Discussion and Summary are organized in Section \ref{sec:discu} and \ref{sec:summary}, respectively.

\section{Data reduction} 
\label{sec:ngc4051_data}

As one of the nearest active galaxies, NGC 4051 has been extensively studied across the X-ray band by numerous space observatories. Its X-ray properties have been investigated by multiple generations of X-ray telescopes, including early missions such as the Einstein Observatory \cite[][]{1983ApJ...269L..31M} and Ginga \cite[][]{1990MNRAS.243..713D}, followed by the Rossi X-Ray Timing Explorer \cite[RXTE,][]{2003MNRAS.338..323L}. More recent observations have been carried out by flagship X-ray facilities including Chandra \cite[][]{2001ApJ...557....2C}, XMM-Newton \cite[][]{2002ApJ...580L.117M}, Suzaku \cite[][]{2009PASJ...61S.299T}, Swift \cite[][]{2013MNRAS.429...75A} and Nuclear Spectroscopic Telescope Array \cite[NuSTAR,][]{2015ApJ...815...66A}.
In this work, to study the X-ray emission of NGC 4051 and its variance with optical/UV bands, we mainly analyzed the data observed by XMM-Newton and Swift telescopes with their simultaneous optical/UV and X-ray observations. 
Moreover, with our primary results in Figure \ref{fig:swift}, NGC 4051 has different optical/UV to X-ray correlations during May-June 2009 with divided by May 27. We only consider the data that were observed from May to June 2009. 


\subsection{Swift}
\label{subsec:swift}

49 Swift observations have well-detected data of NGC 4051 for the XRT (in Photon Counting mode) or UVOT ($UVW1$ band) instruments from May to June 2009. 
Although these Swift observational data have been analyzed and published in \cite{2013MNRAS.429...75A}, rather than extracting data from their figures, we re-analyzed these data. 
We downloaded Swift X-ray data from HEASARC\footnote{\url{https://heasarc.gsfc.nasa.gov/}} and processed these data with the $xrtpipeline$ script. The level 2 event files were filtered in the energy range of $0.3-10.0~\rm keV$ by using the XSELECT, and then used to calculate the source count rate within the source-centered circle of radius 100 arcsec. 

Almost all of the Swift observations in 2009 have UVOT/UVW1 band data, but lack data from other UVOT bands.  
As \cite{2023MNRAS.519..909Z}, we process all Swift optical/UV data with online interactive data analysis tools\footnote{\url{https://swift.ssdc.asi.it/}}. 
The details of the X-ray count rate and $UVW1$ fluxes are displayed in Table \ref{tab:swift}. 

\begin{table}
	\centering
	\caption{The Swift observations of NGC 4051}
	\label{tab:swift}
	\begin{threeparttable}
		\begin{tabular}{l|ccc}
			\hline
			\multicolumn{1}{c}{Obs ID} & \multicolumn{1}{c}{Obs date} & \multicolumn{1}{c}{Rate} & \multicolumn{1}{c}{$f_{\rm \lambda, uvw1}$} \\
			\multicolumn{1}{c}{ } & \multicolumn{1}{c}{ } & \multicolumn{1}{c}{$\rm cts/s$} & \multicolumn{1}{c}{$\rm mJy$} \\
			\multicolumn{1}{c}{(1)} & \multicolumn{1}{c}{(2)} & \multicolumn{1}{c}{(3)} & \multicolumn{1}{c}{(4)} \\
			\hline
			00037585002   &   2009 May 06   &   $0.72 \pm {0.03}$   &   $4.53 \pm {0.02}$  \\  
			00037585003   &   2009 May 07   &   $0.64 \pm {0.02}$   &   $4.67 \pm {0.02}$  \\  
			00037585004   &   2009 May 08   &   $0.94 \pm {0.03}$   &   $4.42 \pm {0.02}$  \\  
			00037585005   &   2009 May 09   &   $0.87 \pm {0.03}$   &   $5.03 \pm {0.02}$  \\  
			00037585006   &   2009 May 10   &   $0.40 \pm {0.02}$   &   $4.47 \pm {0.02}$  \\  
			00037585007   &   2009 May 11   &   $0.58 \pm {0.02}$   &   $4.49 \pm {0.02}$  \\  
			00037585008   &   2009 May 12   &   $0.21 \pm {0.01}$   &   $4.53 \pm {0.02}$  \\  
			00037585009   &   2009 May 13   &   $0.44 \pm {0.02}$   &   $4.39 \pm {0.02}$  \\  
			00037585010   &   2009 May 14   &   $0.32 \pm {0.03}$   &   $4.48 \pm {0.04}$  \\  
			00037585011   &   2009 May 14   &   $0.46 \pm {0.04}$   &   $4.31 \pm {0.04}$  \\  
			00037585012   &   2009 May 14   &   $1.76 \pm {0.06}$   &   $4.90 \pm {0.03}$  \\  
			00037585014   &   2009 May 16   &   $1.03 \pm {0.19}$   &   ...  \\  
			00037585017   &   2009 May 17   &   $0.98 \pm {0.19}$   &   ...  \\  
			00037585020   &   2009 May 19   &   $1.26 \pm {0.16}$   &   ...  \\  
			00037585021   &   2009 May 19   &   $0.95 \pm {0.05}$   &   $5.13 \pm {0.04}$  \\  
			00037585024   &   2009 May 21   &   $1.17 \pm {0.03}$   &   $4.71 \pm {0.02}$  \\  
			00037585025   &   2009 May 22   &   $0.84 \pm {0.03}$   &   $4.57 \pm {0.03}$  \\  
			00037585026   &   2009 May 23   &   $1.42 \pm {0.05}$   &   $4.69 \pm {0.02}$  \\  
			00037585027   &   2009 May 24   &   $0.89 \pm {0.05}$   &   ...  \\  
			00037585028   &   2009 May 25   &   $0.44 \pm {0.03}$   &   $4.47 \pm {0.02}$  \\  
			00037585029   &   2009 May 26   &   $0.63 \pm {0.03}$   &   $4.86 \pm {0.02}$  \\  
			00037585030   &   2009 May 27   &   $0.20 \pm {0.04}$   &   ...  \\  
			00037585031   &   2009 May 28   &   $0.34 \pm {0.03}$   &   $3.92 \pm {0.03}$  \\  
			00037585032   &   2009 May 28   &   $0.19 \pm {0.04}$   &   $4.16 \pm {0.03}$  \\  
			00037585033   &   2009 May 29   &   $0.37 \pm {0.04}$   &   ...  \\  
			00037585036   &   2009 May 31   &   $0.71 \pm {0.03}$   &   $5.01 \pm {0.03}$  \\  
			00037585037   &   2009 Jun 01   &   $0.32 \pm {0.06}$   &   ...  \\  
			00037585038   &   2009 Jun 02   &   $0.29 \pm {0.02}$   &   $5.19 \pm {0.03}$  \\  
			00037585039   &   2009 Jun 03   &   $0.43 \pm {0.03}$   &   $5.45 \pm {0.04}$  \\  
			00037585040   &   2009 Jun 03   &   $0.33 \pm {0.06}$   &   ...  \\  
			00037585041   &   2009 Jun 04   &   $0.81 \pm {0.04}$   &   $5.85 \pm {0.04}$  \\  
			00037585042   &   2009 Jun 04   &   $0.30 \pm {0.10}$   &   ...  \\  
			00037585043   &   2009 Jun 05   &   $0.44 \pm {0.03}$   &   $5.35 \pm {0.04}$  \\  
			00037585044   &   2009 Jun 05   &   $0.25 \pm {0.03}$   &   $5.51 \pm {0.05}$  \\  
			00037585045   &   2009 Jun 06   &   $0.49 \pm {0.03}$   &   $5.59 \pm {0.04}$  \\  
			00037585046   &   2009 Jun 06   &   $0.42 \pm {0.04}$   &   $6.51 \pm {0.06}$  \\  
			00037585049   &   2009 Jun 08   &   $0.31 \pm {0.03}$   &   $4.68 \pm {0.04}$  \\  
			00037585051   &   2009 Jun 09   &   $0.30 \pm {0.03}$   &   $4.78 \pm {0.04}$  \\  
			00037585052   &   2009 Jun 09   &   $0.29 \pm {0.03}$   &   $4.82 \pm {0.04}$  \\  
			00037585053   &   2009 Jun 10   &   $0.32 \pm {0.16}$   &   ...  \\  
			00037585055   &   2009 Jun 11   &   $0.90 \pm {0.04}$   &   ...  \\  
			00037585056   &   2009 Jun 12   &   $0.40 \pm {0.15}$   &   ...  \\  
			00037585057   &   2009 Jun 12   &   $0.31 \pm {0.10}$   &   ...  \\  
			00037585058   &   2009 Jun 13   &   $0.60 \pm {0.05}$   &   ...  \\  
			00037585059   &   2009 Jun 13   &   $0.80 \pm {0.04}$   &   ...  \\  
			00037585060   &   2009 Jun 14   &   $1.51 \pm {0.22}$   &   ...  \\  
			00037585061   &   2009 Jun 15   &   $0.55 \pm {0.03}$   &   $5.66 \pm {0.03}$  \\  
			00037585062   &   2009 Jun 16   &   $0.68 \pm {0.14}$   &   ...  \\  
			00037585063   &   2009 Jun 16   &   $1.06 \pm {0.12}$   &   ...  \\  
			\hline
		\end{tabular}
		\begin{tablenotes}
			\footnotesize
			\item In this Table, Column (1): Swift observational ID; Column (2): Swift observational date; Column (3): X-ray count rate in $0.3-10.0~\rm keV$; Column (4): UV flux in UVW1 band.
		\end{tablenotes}
	\end{threeparttable}
\end{table}%

\begin{figure*}
	\centering
	\includegraphics[width=2\columnwidth]{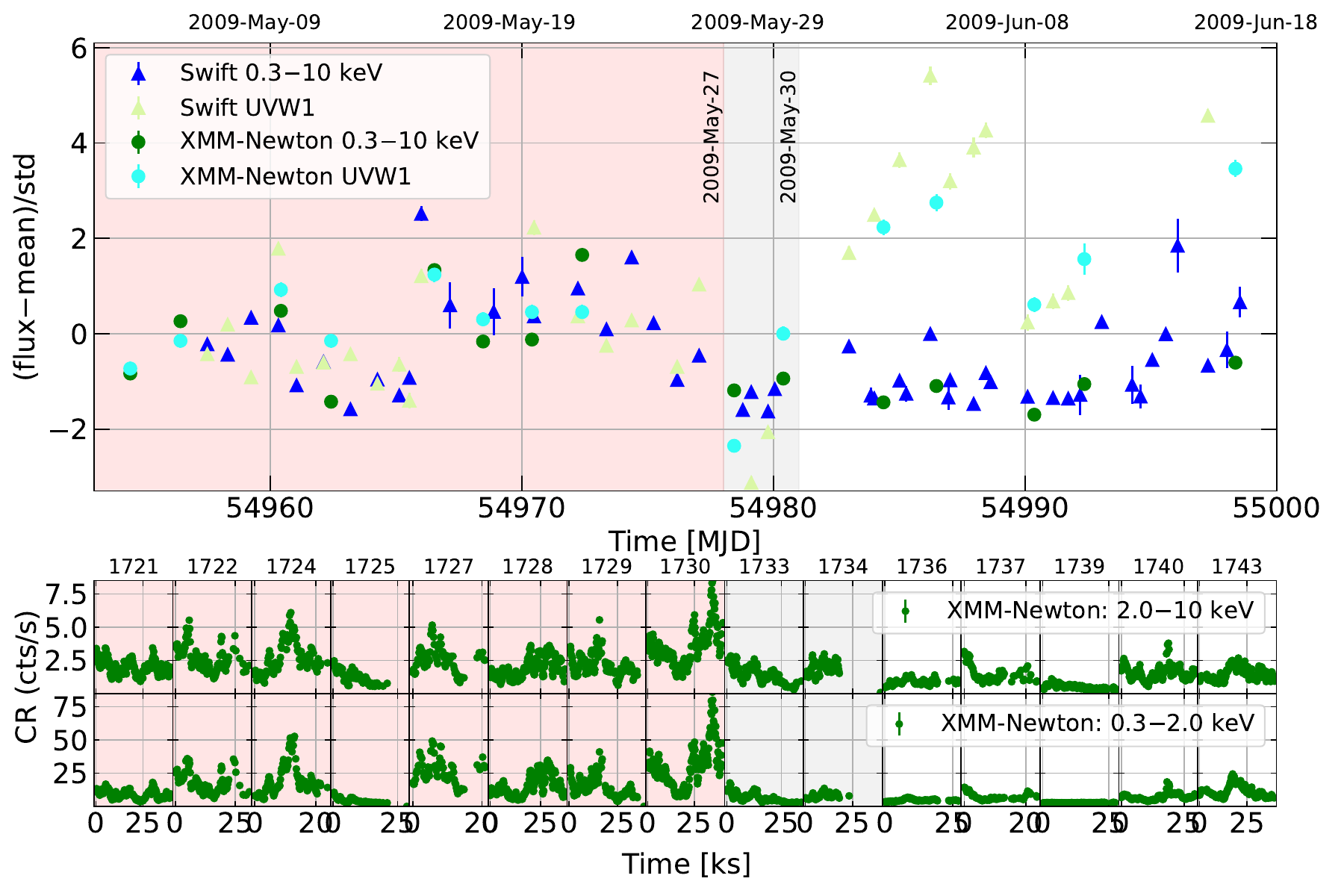}
	\caption{The light curves of NGC 4051. (Top panel) Long-term optical/UV and $0.3-10\,\rm keV$ X-ray light curves of Swift and XMM-Newton data observed between May and June 2009. (Middle and bottom panels) Short-term soft and hard X-ray light curves of each XMM-Newton observation. The red-, gray-shaded, and white regions correspond to different optical/UV-X-ray correlation periods.}
	\label{fig:swift}
\end{figure*}

A comparative analysis of the optical and X-ray light curves from Figure 2 in \cite{2013MNRAS.429...75A} revealed temporally distinct variability patterns in NGC 4051 during May-June 2009, divided by the May 27 transition epoch. 
Such results are also approved by the Swift observations in Table \ref{tab:swift}. 
Specifically, the post-May 27 interval exhibited stronger optical flux variations ($F_{\rm var,uvw1}=9.56\pm0.22 \%$) concurrent suppressed X-ray fluctuations ($F_{\rm var,0.3-10 keV}=34.97\pm3.87 \%$) compared to pre-transition levels ($F_{\rm var,uvw1}=4.85\pm0.14 \%$, $F_{\rm var,0.3-10 keV}=46.14\pm2.03 \%$), demonstrating different physical emission process. 
Where $F_{\rm var}=\sqrt{\frac{\sigma^2-<\sigma_{err}^2>}{<x>^2})}$ is the fractional variability \cite[e.g.,][]{2010MNRAS.403..605B} for $N$ measurement data $x_i$ (with error $\sigma_i$, mean value $<x>$), $\sigma$ is the flux variance ($\sigma^2=\frac{1}{N}\sum(x_i - <x>^2)$), $<\sigma_{err}^2>=\frac{1}{N}\sum \sigma_i^2$ is the mean of the squared measurement errors. 
The error on $F_{\rm var}$ is given by $F_{\rm var, err} = \sqrt{(\sqrt{\frac{1}{2N}} \frac{<\sigma_{err}^2>}{<x>^2 F_{var}})^2+(\sqrt{\frac{<\sigma_{err}^2>}{N}} \frac{1}{<x>})^2}$. 
Considering the lower fractional variability of the optical/UV band than that of the X-ray band \cite[e.g.,][]{2013MNRAS.429...75A}, we plot Swift UVW1 and X-ray band ($0.3-10\,\rm keV$) light curves in Figure \ref{fig:swift} upper panel as \cite{2003MNRAS.343.1341S} with renormalized by their respective standard deviation (after mean subtraction) of the measurements before May 27, 2009. 

\subsection{XMM-Newton}
\label{subsec:xmm}

NGC 4051 has been observed by the XMM-Newton telescope on both optical/UV and X-ray bands 20 times from 2001 to 2023, with 15 of these observations being simultaneous with those by the Swift telescope during May to June 2009. 
The XMM-Newton observations are listed in the first four columns of Table \ref{tab:xmm}. 
\cite{2011MNRAS.413.2489V} presents an analysis of the high-frequency X-ray variability of NGC 4051 based on a series of XMM-Newton observations taken in 2009 with the power spectral distribution. 
Combined with XMM-Newton and Swift simultaneous UV to X-ray data, \cite{2013MNRAS.429...75A} analyzed NGC 4051 UV to X-ray variability over 45 days in 2009. 
With spectral analysis for XMM-Newton, Suzaku, and PCA/RXTE observations, \cite{2018A&A...613A..48S} found a tight correlation between X-ray photon index and accretion rate. 
To study the mechanisms of softer-when-brighter properties in primary power-law X-ray spectra of AGNs at moderate to high accretion rates, \cite{2020SCPMA..6329512W} scrutinized the power-law spectral variability of NGC 4051 with excluded photons $<2\, \rm keV$ to avoid contamination from the soft excess. 
In summary, almost all the XMM-Newton observations of NGC 4051 have been investigated by spectral fitting and light curve analysis with different methods or passbands. Instead of directly taking the results from the literature, to avoid or reduce the systematic difference for each observation reduction, we decided to re-analyze all the XMM-Newton data for NGC 4051 uniformly.

\begin{table*}
	\centering
	\caption{The XMM-Newton observations of NGC 4051}
	\label{tab:xmm}
	\begin{threeparttable}
		\begin{tabular}{l|crcrcccc}
			\hline
			\multicolumn{1}{c}{Obs ID} & \multicolumn{1}{c}{Rev.} & \multicolumn{1}{c}{Obs date} & \multicolumn{1}{c}{Exp.} & \multicolumn{1}{c}{pn} & \multicolumn{1}{c}{UVW1} & \multicolumn{1}{c}{$F_{0.3-10~\rm keV}$} & \multicolumn{1}{c}{$F_{0.3-2.0~\rm keV}$} & \multicolumn{1}{c}{$F_{2.0-10~\rm keV}$}  \\
			\multicolumn{2}{c}{ } & \multicolumn{1}{c}{} & \multicolumn{1}{c}{$\rm ks$} & \multicolumn{1}{c}{$\rm cts/s$} & \multicolumn{1}{c}{$\rm Mag$} & \multicolumn{3}{c}{} \\
			\multicolumn{1}{c}{(1)} & \multicolumn{1}{c}{(2)} & \multicolumn{1}{c}{(3)} & \multicolumn{1}{c}{(4)} & \multicolumn{1}{c}{(5)} & \multicolumn{1}{c}{(6)} & \multicolumn{1}{c}{(7)} & \multicolumn{1}{c}{(8)} & \multicolumn{1}{c}{(9)} \\
			\hline
			0606320101 & 1721 & 2009-05-03 & 45.717 & 11.21$\pm$0.02 & 13.36$\pm$0.01 & 0.335$\pm$0.002 & 0.354$\pm$0.002 & 0.271$\pm$0.005 \\
			0606320201 & 1722 & 2009-05-05 & 45.645 & 21.05$\pm$0.04 & 13.32$\pm$0.01 & 0.430$\pm$0.002 & 0.432$\pm$0.002 & 0.343$\pm$0.005 \\
			0606320301 & 1724 & 2009-05-09 & 45.548 & 23.00$\pm$0.04 & 13.25$\pm$0.01 & 0.580$\pm$0.002 & 0.602$\pm$0.002 & 0.436$\pm$0.006 \\
			0606320401 & 1725 & 2009-05-11 & 45.447 & 5.90$\pm$0.02 & 13.32$\pm$0.01 & 0.412$\pm$0.004 & 0.418$\pm$0.005 & 0.393$\pm$0.009 \\
			0606321301 & 1727 & 2009-05-15 & 32.644 & 30.68$\pm$0.06 & 13.23$\pm$0.01 & 0.359$\pm$0.002 & 0.360$\pm$0.002 & 0.364$\pm$0.007 \\
			0606321401 & 1728 & 2009-05-17 & 42.433 & 17.23$\pm$0.03 & 13.29$\pm$0.01 & 0.491$\pm$0.002 & 0.509$\pm$0.002 & 0.380$\pm$0.006 \\
			0606321501 & 1729 & 2009-05-19 & 41.813 & 17.60$\pm$0.03 & 13.28$\pm$0.01 & 0.492$\pm$0.002 & 0.511$\pm$0.002 & 0.362$\pm$0.006 \\
			0606321601 & 1730 & 2009-05-21 & 41.936 & 33.53$\pm$0.04 & 13.28$\pm$0.01 & 0.537$\pm$0.001 & 0.554$\pm$0.001 & 0.405$\pm$0.004 \\
			\hline
			0606321701 & 1733 & 2009-05-27 & 44.919 & 8.03$\pm$0.02 & 13.48$\pm$0.01 & 0.497$\pm$0.003 & 0.517$\pm$0.003 & 0.429$\pm$0.007 \\
			0606321801 & 1734 & 2009-05-29 & 43.726 & 10.26$\pm$0.03 & 13.31$\pm$0.01 & 0.250$\pm$0.003 & 0.226$\pm$0.004 & 0.260$\pm$0.007 \\
			0606321901 & 1736 & 2009-06-02 & 44.891 & 5.78$\pm$0.02 & 13.17$\pm$0.01 & 0.176$\pm$0.004 & 0.171$\pm$0.004 & 0.228$\pm$0.011 \\
			0606322001 & 1737 & 2009-06-04 & 39.756 & 8.86$\pm$0.03 & 13.14$\pm$0.01 & 0.347$\pm$0.004 & 0.342$\pm$0.004 & 0.398$\pm$0.010 \\
			0606322101 & 1739 & 2009-06-08 & 43.545 & 3.46$\pm$0.01 & 13.27$\pm$0.01 & 0.097$\pm$0.005 & 0.015$\pm$0.014 & 0.311$\pm$0.014 \\
			0606322201 & 1740 & 2009-06-10 & 44.453 & 9.22$\pm$0.02 & 13.21$\pm$0.02 & 0.317$\pm$0.003 & 0.325$\pm$0.003 & 0.349$\pm$0.006 \\
			0606322301 & 1743 & 2009-06-16 & 42.717 & 13.25$\pm$0.02 & 13.10$\pm$0.01 & 0.384$\pm$0.002 & 0.400$\pm$0.002 & 0.295$\pm$0.006 \\
			\hline
		\end{tabular}
		\begin{tablenotes}
			\footnotesize
			\item In this Table, Column (1): XMM-Newton observational ID; Column (2): Revolution number; Column (3): XMM-Newton observational date; Column (4): Exposure time; Column (5): X-ray count rate in $0.3-10.0~\rm keV$; Column (6): UV flux in UVW1 band; Column (7-9): Short-term X-ray fractional variability for each data.
		\end{tablenotes}
	\end{threeparttable}
\end{table*}%

All XMM-Newton observations of NGC 4051 have both pn and MOS detection. In this work, we use only the EPIC pn data for the higher spectral resolution than that of the MOS data. 
We processed all data with XMM-Newton Scientific Analysis Software (SAS) following the steps in The XMM-Newton ABC Guide\footnote{\url{https://heasarc.gsfc.nasa.gov/docs/xmm/abc/}}. 
The pn data were reprocessed with the $epproc$ script in SAS-18.0.0 and filtered with the same standard filters as \cite{2023MNRAS.519..909Z}: PATTERN in the range of $0-4$, energy in the range of $0.2-15.0~\rm keV$, and \#XMMEA\_EP. 
The large flare time intervals were filtered out with an examination of the light curve. 
The source spectra of NGC 4051 were then extracted from the source-centered radius of 32 arcsec for all data, with the background source-free region of 40 arcsec around the object. 
The redistribution matrix files and ancillary response files were created with the $rmfgen$ and $arfgen$ scripts. 
All X-ray spectra are not or less affected by pile-up as checked by $epatplot$, and rebinned with a minimum of 20 counts for the background-subtracted spectral channel, and oversampled the intrinsic energy resolution by a factor no larger than three. 
Our source count rates in $0.3-10\, \rm keV$ are shown in Table \ref{tab:xmm} Column (5).

The short-term source X-ray light curves were created from the pn cleaned data, and filtered for particle background event files. The source and background light curves were extracted with the $evselect$ using the same regions for spectral extraction. 
All light curves for different energy ranges ($0.3-2.0~\rm keV$, $2.0-10~\rm keV$, $2.0-4.5~\rm keV$, $4.5-10~\rm keV$, and $0.3-10~\rm keV$) were corrected with the background subtraction by the $epiclccorr$ tool, using a uniform time bin size of 500 $\rm s$. 
The soft and hard X-ray light curves are shown in Figure \ref{fig:swift} bottom panels, and the correlated fractional variability are presented in Table \ref{tab:xmm}. 

Optical/UV OM image data have been fully processed by the XMM-SAS Pipeline. We directly use the pipeline results to calculate the optical flux in each band. To gain the photon count rate or Magnitude, the interactive tool $omsource$ has been used. All corrected OM/UVW1 magnitudes are shown in Table \ref{tab:xmm} Column (6). 
The UVW1 magnitudes and XRT count rates were also added to Figure \ref{fig:swift} upper panel with the same method as the Swift data. 

The soft X-ray data from the reflection grating spectrometers \cite[RGS,][]{2001A&A...365L...7D} had been studied by \cite{2011MNRAS.413.1251P, 2011MNRAS.415.2379P} with integrated spectra at representative high and low flux levels. 
To investigate the soft X-ray features of individual observation, we reprocess the RGS data with the $rgsproc$ script by excluding the high background interval.

\section{Results}
\label{Sec:results}

\subsection{light curve}
\label{subsec:lc}

As shown in Figure \ref{fig:swift}, both X-ray and UV light curves of NGC 4051 show variations on long and short time scales. 
The long-term optical/UV to X-ray light curves of Swift and XMM-Newton observations had been analyzed in \cite{2013MNRAS.429...75A}. 
During the period of May-June 2009, contrasting with the results from \cite{2013MNRAS.429...75A}, we discovered that NGC 4051 exhibited strong optical/UV-X-ray variability correlations in Figure \ref{fig:swift} upper panel before 27 May 2009 with Sift data analysis of the Spearman coefficient $r_{\rm s}=0.63$ at $p=0.00664$ confidence level, but no significant correlations were observed after 30 May with the Spearman test of $p=0.16667$. The interval from 27 to 30 May likely represents a transitional phase. Figure \ref{fig:swift} demarcates these three epochs using distinct background colors. 
The XMM-Newton observation data further support the Swift light curves correlation results. 
Interestingly, during the UV outburst of NGC 4051 after May 27, 2009, its X-ray maintained a relatively stable luminosity level, suggesting a potential physical connection between the variations in the two bands. Further discussion is provided in Section \ref{subsec:PossibleExp}.

About the XMM-Newton short-term X-ray light curves in Figure \ref{fig:swift} middle and bottom panels, we observed distinct variability behaviors in the short-term soft X-ray emission of NGC 4051 before and after May 27. Specifically, the amplitude of both soft and hard X-ray variability exhibited a marked decrease following May 27. 
As documented in the last two columns of Table \ref{tab:xmm}, both soft and hard X-ray fractional variability values in most post-May 27 XMM-Newton observations exhibited an overall reduction compared to pre-event measurements (with the exception of dataset 000606321701 with revolution 1733). 
The dataset 1733 has soft and hard X-ray fractional variability comparable to those of the previous observation period. 

\begin{figure}
	\centering
	\includegraphics[width=1\columnwidth]{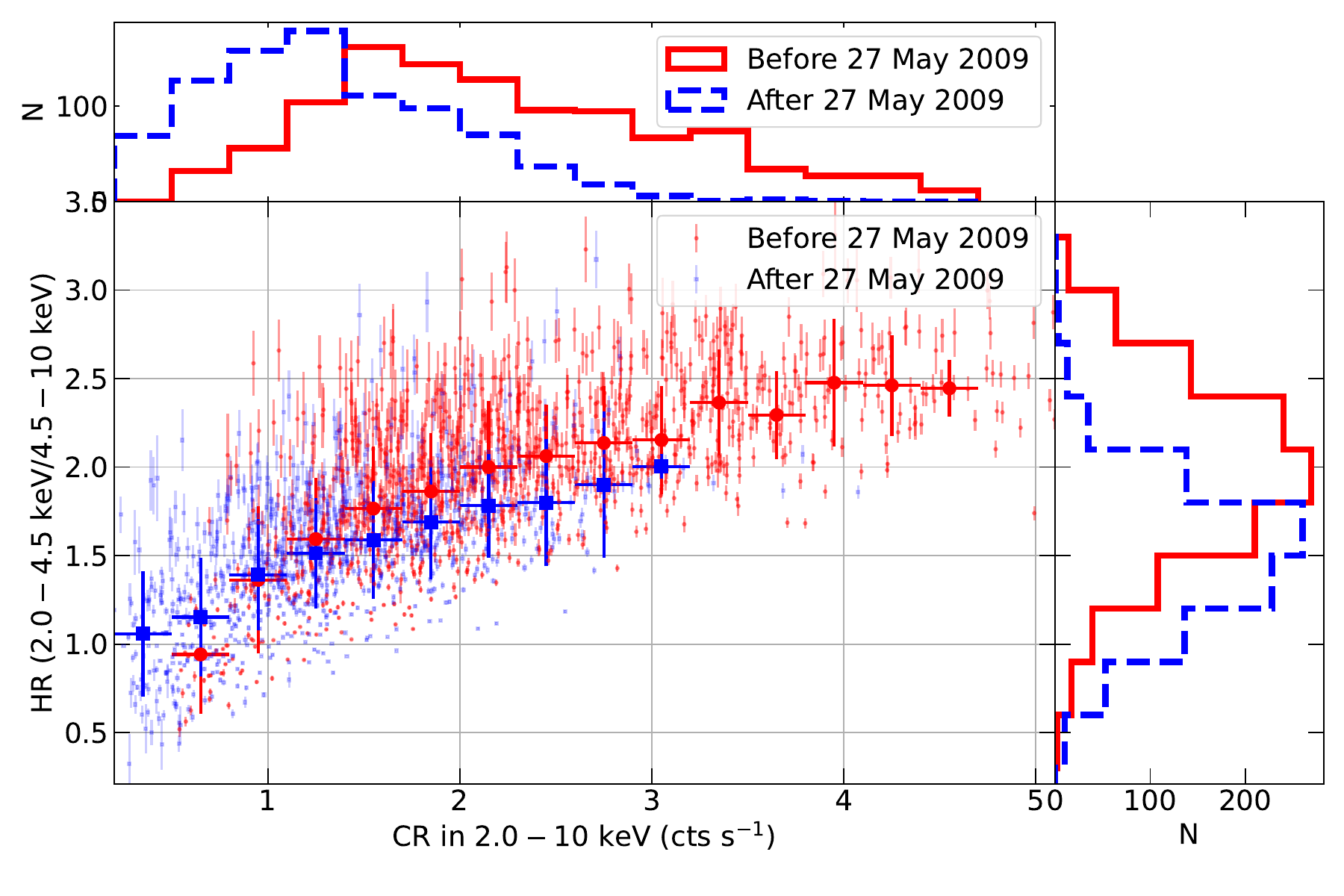}
	\caption{The hardness ratio versus X-ray luminosity correlation in NGC 4051's short-term hard X-ray light curves. Red and blue data points represent X-ray observations before and after May 27, 2009, respectively. Larger solid circular markers indicate median hardness ratios within each count-rate bin. The top and right 2D histograms show the distribution of hard X-ray photon count rate and hardness ratio, respectively.}
	\label{fig:xmm-HR}
\end{figure}

The presence of differences in the hard X-ray flux variations of NGC 4051 before and after May 27, as shown in Figure \ref{fig:swift} middle panels, prompts us to further investigate the relationship between the hard X-ray hardness ratio and luminosity variations. 
Figure \ref{fig:xmm-HR} displays the correlation between hard X-ray hardness ratio (defined as $R = CR_{2.0-4.5\,\rm keV}/CR_{4.5-10\,\rm keV}$) and hard X-ray photon count rate, along with the distribution of these two parameters. 
Statistically significant differences were observed in Figure \ref{fig:xmm-HR} between parameter distributions: Compared to pre-May 27th measurements, the hard X-ray luminosity showed overall dimming (KS test: $D = 0.366$, $P = \rm 1.7567E-67$), accompanied by lower hardness ratios (KS test: $D = 0.306$, $P = \rm 4.2041E-47$). 
However, when considering error distributions, the hardness ratio's variation with luminosity (photon count rate) during both pre- and post-May 27th periods generally followed the same trend, manifesting as brightening with spectral softening. 

\subsection{Spectral fitting}

The historical X-ray data of NGC 4051 have been largely analyzed with spectral modeling by other authors. 
\cite{2001ApJ...557....2C} found that NGC 4051 has been absorbed by several outflows with Chandra HETG data, and revealed that the soft excess can be fitted with a blackbody model. 
The complex X-ray spectral behaviour exhibited in NGC 4051 during its November 2002 low-flux state can be explained through a combination of ionised absorption, thermal components, and reflection processes \cite[][]{2004MNRAS.347.1345U}. 
A comparative analysis of the XMM-Newton observations from 2001 and 2002 reveals consistent Fe K$\alpha$ line fluxes in NGC 4051 across both high- and low-flux states, indicating similar reflection components in the respective X-ray spectra \cite[][]{2004MNRAS.350...10P}. \cite{2004MNRAS.350...10P} interpreted the major differences between the high- and low-flux hard X-ray spectra in terms of the varying ionization (opacity) of a substantial column of outflowing gas.
Making use of both RGS and EPIC-pn data in 2009, \cite{2016A&A...596A..79S} performed a detailed analysis through a time-dependent photoionization code in combination with spectral and Fourier spectral-timing techniques, and found that the absorbing gas can likewise produce a time delay between the broader soft and hard bands. 
The rms spectra of NGC 4051 have a sharp peak and several dips, which can be explained by variable absorption features and non-variable emission lines. \cite{2017MNRAS.466.3259M} proposed that NGC 4051 has two types of warm absorber outflows. 
With X-ray spectral fitting, \cite{2018A&A...613A..48S} presented the evolution of the spectral (photon) index from lower to higher accretion rate. 

Although \cite{2018A&A...613A..48S} Table A.4 presents the complete 2009 observational spectral fitting results for NGC 4051, their adopted model (gabs$\times$(BMC+Laor+redge)) neglected the effects of outflow absorption \cite[][]{2016A&A...596A..79S, 2017MNRAS.466.3259M} and reflection components \cite[][]{2006MNRAS.368..903P}. Consequently, the photon indexes measured during the low-flux state after May 27 exhibit unusually flat values ($\Gamma \sim 1.35$), significantly lower than those typically observed in AGN X-ray spectra \cite[$\Gamma \sim 2.0$, ][]{2021RAA....21....4Z}. 

\begin{figure}
	\centering
	\includegraphics[width=1\columnwidth]{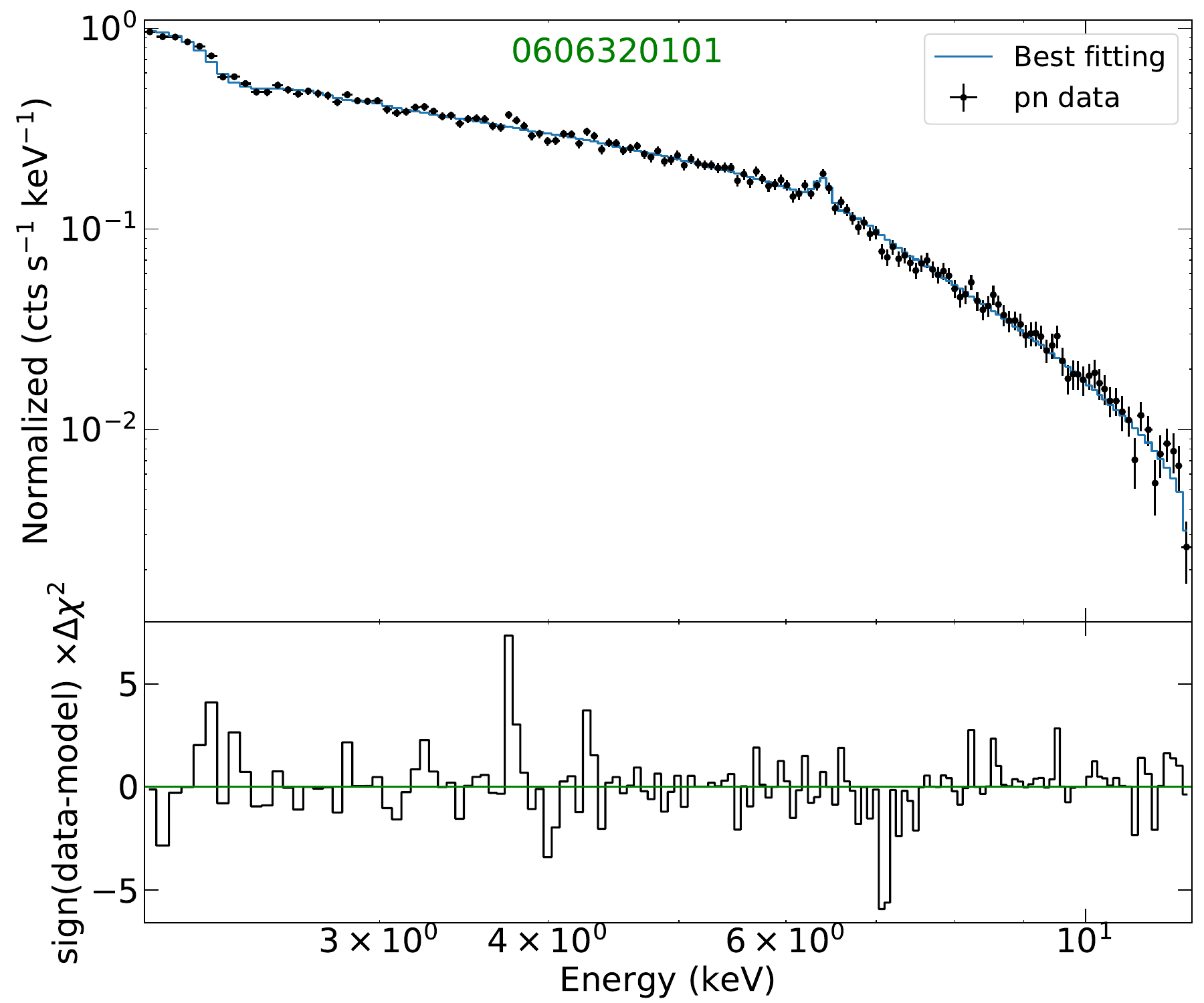}
	\caption{The example of the XMM-Newton hard X-ray spectrum fitting of NGC 4051.}
	\label{fig:xmm-hard_spec}
\end{figure}

To investigate on the spectral component evolution with the UV flare in 2009, in this work, we fit all XMM-Newton X-ray spectra uniformly with XSPEC 12.15.0. 
For the historical results about ionized absorption and reflection \cite[e.g.,][]{2004MNRAS.350...10P}, we first modeled the $2.0-12.0\ \rm keV$ X-ray spectra with the Galaxy corrected partially ionized absorbed reflection plus Gaussian model ($phabs \times zxipcf \times (relxill + zgauss)$). 
For these input parameters, the H {\footnotesize I} column density \cite[$nH = 1.19 \times 10^{20}\ \rm cm^{-2}$,][]{2016A&A...594A.116H} and redshift $z=0.002336$ were fixed. 
Under the default configuration, our analysis revealed that the black hole spin values consistently yielded values of approximately 0.998 (approaching the theoretical maximum), while the fitted values of the spin of black holes and the iron abundance from different observations show significant differences, which is contrary to common sense. 
In the following fitting, based on our initial fitting results, we fix the black hole spin, the inclination angle, and the iron abundance of the $relxill$ \cite[][]{2014ApJ...782...76G, 2014MNRAS.444L.100D} to $a=0.998$, $incl=30\ \rm deg$, and $A_{\rm Fe}=0.5$. 
Our ionized absorbed reflection model can well fit all XMM-Newton hard X-ray spectra. 
Figure \ref{fig:xmm-hard_spec} shows an example of the XMM-Newton hard X-ray spectrum (Observation ID 0606320101) fitting of NGC 4051.  

\begin{landscape}
	\begin{table}
		\tiny
		\caption{The spectral fitting results for XMM-Newton observations of NGC 4051}
		\label{tab:xmm_fitting}
		\begin{threeparttable}
			\begin{tabular}{l|lcccccccccccccccccccccc}
				\hline
				\multicolumn{1}{c}{} & \multicolumn{3}{c}{Zxipcf} && \multicolumn{2}{c}{Zedge} && \multicolumn{2}{c}{bbody} && \multicolumn{3}{c}{Zgauss} && \multicolumn{3}{c}{Zgauss} && \multicolumn{4}{c}{Relxill} & \multicolumn{1}{c}{ } \\
				\cline{2-4} \cline{6-7} \cline{9-10} \cline{12-14} \cline{16-18} \cline{20-23} \\
				\multicolumn{1}{c}{Obs ID} & \multicolumn{1}{c}{$nH$} & \multicolumn{1}{c}{$\log \xi$} & \multicolumn{1}{c}{$CF$} && \multicolumn{1}{c}{$E_{\rm edge}$} & \multicolumn{1}{c}{$\tau_{\rm max}$} && \multicolumn{1}{c}{$kT$} & \multicolumn{1}{c}{$Norm$} && \multicolumn{1}{c}{$E$} & \multicolumn{1}{c}{$\sigma$} & \multicolumn{1}{c}{$Norm$} && \multicolumn{1}{c}{$E$} & \multicolumn{1}{c}{$\sigma$} & \multicolumn{1}{c}{$Norm$} && \multicolumn{1}{c}{$\Gamma$} & \multicolumn{1}{c}{$\log \xi$} & \multicolumn{1}{c}{$refl_{\rm frac}$} & \multicolumn{1}{c}{$Norm$} & \multicolumn{1}{c}{$\chi^2$/dof} \\ 
				\multicolumn{1}{c}{ } & \multicolumn{1}{c}{$10^{22}\,\rm cm^{-2}$} & \multicolumn{2}{c}{} && \multicolumn{1}{c}{$\rm keV$} & \multicolumn{1}{c}{} && \multicolumn{1}{c}{$\rm keV$} & \multicolumn{1}{c}{$E-04$} && \multicolumn{1}{c}{$\rm keV$} & \multicolumn{1}{c}{$\rm keV$} & \multicolumn{1}{c}{$10^{-4}$} && \multicolumn{1}{c}{$\rm keV$} & \multicolumn{1}{c}{$\rm keV$} & \multicolumn{1}{c}{$10^{-5}$} && \multicolumn{1}{c}{ } & \multicolumn{1}{c}{ } & \multicolumn{1}{c}{ } & \multicolumn{1}{c}{$10^{-5}$} & \multicolumn{1}{c}{ } \\ 
				\multicolumn{1}{c}{(1)} & \multicolumn{1}{c}{(2)} & \multicolumn{1}{c}{(3)} & \multicolumn{1}{c}{(4)} && \multicolumn{1}{c}{(5)} & \multicolumn{1}{c}{(6)} && \multicolumn{1}{c}{(5)} & \multicolumn{1}{c}{(6)}  && \multicolumn{1}{c}{(7)} & \multicolumn{1}{c}{(8)} & \multicolumn{1}{c}{(9)} && \multicolumn{1}{c}{(10)} & \multicolumn{1}{c}{(11)} & \multicolumn{1}{c}{(12)} && \multicolumn{1}{c}{(13)} & \multicolumn{1}{c}{(14)} & \multicolumn{1}{c}{(15)} & \multicolumn{1}{c}{(16)} & \multicolumn{1}{c}{(17)} \\
				\hline
				0606320101 & $3.67_{-0.21}^{+0.23}$ & $2.09_{-0.06}^{+0.07}$ & $>0.85$ & & $0.60_{-0.01}^{+0.01}$ & $0.51_{-0.15}^{+0.16}$ & & $0.136_{-0.008}^{+0.009}$ & $0.95_{-0.06}^{+0.06}$ & & $0.894_{-0.008}^{+0.007}$ & $0.03_{-0.01}^{+0.01}$ & $5.75_{-1.67}^{+2.14}$ & & $6.411_{-0.020}^{+0.033}$ & $<0.01$ & $1.43_{-0.35}^{+0.36}$ & & $1.75_{-0.16}^{+0.16}$ & $3.20_{-0.15}^{+0.07}$ & $3.09_{-1.44}^{+5.73}$ & $3.95_{-2.36}^{+2.42}$ & 187/174 \\
				0606320201 & $1.35_{-0.08}^{+0.11}$ & $1.25_{-0.15}^{+0.14}$ & $0.58_{-0.03}^{+0.03}$ & & $0.91_{-0.01}^{+0.02}$ & $0.84_{-0.14}^{+0.16}$ & & $0.120_{-0.003}^{+0.003}$ & $3.15_{-0.18}^{+0.17}$ & & $-$ & $-$ & $-$ & & $6.371_{-0.050}^{+0.049}$ & $<0.10$ & $1.55_{-0.68}^{+0.75}$ & & $1.82_{-0.11}^{+0.16}$ & $3.28_{-0.08}^{+0.04}$ & $6.68_{-2.00}^{+10.67}$ & $2.26_{-0.36}^{+2.33}$ & 208.51/170 \\
				0606320301 & $9.30_{-1.43}^{+15.01}$ & $3.49_{-0.04}^{+0.03}$ & $>0.64$ & & $0.89_{-0.01}^{+0.01}$ & $1.25_{-0.20}^{+0.22}$ & & $0.118_{-0.002}^{+0.002}$ & $3.80_{-0.07}^{+0.07}$ & & $-$ & $-$ & $-$ & & $6.467_{-0.047}^{+0.048}$ & $0.09_{-0.06}^{+0.05}$ & $2.38_{-0.70}^{+0.78}$ & & $1.73_{-0.48}^{+0.12}$ & $>2.53$ & $1.98_{-0.31}^{+3.81}$ & $3.24_{-0.02}^{+2.07}$ & 225.73/169 \\
				0606320401 & $12.54_{-0.57}^{+0.62}$ & $2.01_{-0.07}^{+0.06}$ & $0.83_{-0.03}^{+0.03}$ & & $0.74_{-0.01}^{+0.01}$ & $>1.39$ & & $0.135_{-0.006}^{+0.007}$ & $1.51_{-0.18}^{+0.20}$ & & $0.889_{-0.014}^{+0.010}$ & $0.06_{-0.02}^{+0.01}$ & $9.05_{-3.12}^{+2.21}$ & & $6.390_{-0.107}^{+0.032}$ & $<0.08$ & $2.06_{-0.52}^{+0.54}$ & & $1.66_{-0.17}^{+0.23}$ & $3.76_{-0.05}^{+0.07}$ & $3.50_{-1.38}^{+4.44}$ & $2.28_{-1.41}^{+8.62}$ & 197.12/156 \\
				0606321301 & $7.50_{-0.78}^{+1.13}$ & $2.00_{-0.26}^{+0.15}$ & $0.42_{-0.05}^{+0.05}$ & & $0.92_{-0.02}^{+0.04}$ & $1.42_{-0.29}^{+0.36}$ & & $0.123_{-0.003}^{+0.003}$ & $3.58_{-0.22}^{+0.34}$ & & $-$ & $-$ & $-$ & & $6.420_{-0.048}^{+0.042}$ & $0.04_{-0.03}^{+0.17}$ & $2.12_{-0.73}^{+1.49}$ & & $2.08_{-0.11}^{+0.16}$ & $3.22_{-0.03}^{+0.06}$ & $6.26_{-1.96}^{+5.23}$ & $2.80_{-1.00}^{+1.67}$ & 166.82/160 \\
				0606321401 & $89.84_{-29.26}^{+148.18}$ & $4.21_{-0.18}^{+0.10}$ & $>0.66$ & & $0.91_{-0.02}^{+0.02}$ & $0.52_{-0.14}^{+0.14}$ & & $0.107_{-0.001}^{+0.001}$ & $2.38_{-0.02}^{+0.02}$ & & $-$ & $-$ & $-$ & & $6.378_{-0.038}^{+0.040}$ & $0.12_{-0.05}^{+0.06}$ & $2.45_{-0.56}^{+0.62}$ & & $1.63_{-0.18}^{+0.22}$ & $3.20_{-0.10}^{+0.06}$ & $1.90_{-0.56}^{+0.99}$ & $5.24_{-1.65}^{+1.68}$ & 299.82/170 \\
				0606321501 & $1.45_{-0.22}^{+0.11}$ & $-0.55_{-0.71}^{+0.19}$ & $0.47_{-0.02}^{+0.03}$ & & $0.76_{-0.01}^{+0.01}$ & $0.50_{-0.08}^{+0.07}$ & & $0.110_{-0.001}^{+0.002}$ & $4.17_{-0.19}^{+0.32}$ & & $-$ & $-$ & $-$ & & $6.398_{-0.040}^{+0.040}$ & $0.10_{-0.05}^{+0.05}$ & $2.29_{-0.58}^{+0.64}$ & & $1.87_{-0.07}^{+0.12}$ & $3.10_{-0.06}^{+0.03}$ & $3.23_{-0.90}^{+1.58}$ & $3.92_{-0.11}^{+1.30}$ & 285.82/175 \\
				0606321601 & $108.89_{-10.60}^{+9.94}$ & $2.73_{-0.16}^{+0.14}$ & $0.25_{-0.01}^{+0.04}$ & & $0.89_{-0.01}^{+0.01}$ & $1.14_{-0.12}^{+0.12}$ & & $0.115_{-0.001}^{+0.001}$ & $6.06_{-0.17}^{+0.17}$ & & $-$ & $-$ & $-$ & & $6.409_{-0.032}^{+0.031}$ & $<0.10$ & $2.52_{-0.67}^{+0.67}$ & & $1.83_{-0.21}^{+0.32}$ & $>4.16$ & $2.59_{-1.42}^{+13.96}$ & $4.24_{-0.17}^{+7.89}$ & 271.06/174 \\
				\hline
				0606321701 & $3.10_{-0.40}^{+0.44}$ & $0.60_{-0.21}^{+0.32}$ & $0.53_{-0.02}^{+0.02}$ & & $0.73_{-0.01}^{+0.01}$ & $1.93_{-0.42}^{+0.65}$ & & $0.111_{-0.003}^{+0.004}$ & $1.74_{-0.07}^{+0.13}$ & & $0.869_{-0.011}^{+0.012}$ & $<0.01$ & $0.99_{-0.34}^{+0.33}$ & & $6.405_{-0.030}^{+0.030}$ & $<0.09$ & $1.38_{-0.34}^{+0.35}$ & & $1.72_{-0.10}^{+0.21}$ & $3.26_{-0.28}^{+0.03}$ & $3.24_{-1.44}^{+3.14}$ & $2.66_{-2.83}^{+2.04}$ & 191.63/168 \\
				0606321801 & $3.63_{-0.26}^{+0.42}$ & $-0.65_{-0.14}^{+0.33}$ & $0.84_{-0.01}^{+0.01}$ & & $0.74_{-0.01}^{+0.01}$ & $>2.74$ & & $0.120_{-0.003}^{+0.003}$ & $5.94_{-0.54}^{+0.74}$ & & $0.888_{-0.014}^{+0.015}$ & $<0.07$ & $6.33_{-2.37}^{+2.68}$ & & $6.399_{-0.070}^{+0.061}$ & $<0.17$ & $1.83_{-0.81}^{+0.95}$ & & $2.16_{-0.17}^{+0.22}$ & $3.24_{-0.09}^{+0.07}$ & $5.69_{-2.92}^{+6.68}$ & $3.58_{-1.51}^{+2.94}$ & 204.17/161 \\
				0606321901 & $14.68_{-0.56}^{+0.69}$ & $1.20_{-0.06}^{+0.08}$ & $0.87_{-0.01}^{+0.01}$ & & $0.73_{-0.02}^{+0.02}$ & $1.22_{-0.32}^{+0.56}$ & & $0.122_{-0.008}^{+0.010}$ & $3.34_{-0.39}^{+0.44}$ & & $0.901_{-0.010}^{+0.011}$ & $<0.04$ & $4.66_{-1.19}^{+2.41}$ & & $6.447_{-0.040}^{+0.041}$ & $<0.12$ & $1.97_{-0.65}^{+0.74}$ & & $2.23_{-0.39}^{+0.36}$ & $3.42_{-0.20}^{+0.14}$ & $4.55_{-1.98}^{+8.29}$ & $2.73_{-1.38}^{+3.51}$ & 188.29/150 \\
				0606322001 & $5.70_{-0.77}^{+0.47}$ & $1.92_{-0.20}^{+0.07}$ & $0.79_{-0.02}^{+0.08}$ & & $0.62_{-0.01}^{+0.03}$ & $0.30_{-0.11}^{+0.11}$ & & $0.150_{-0.012}^{+0.010}$ & $1.38_{-0.11}^{+0.40}$ & & $0.896_{-0.011}^{+0.010}$ & $<0.03$ & $3.81_{-1.01}^{+2.56}$ & & $6.455_{-0.051}^{+0.052}$ & $<0.15$ & $2.20_{-0.72}^{+0.79}$ & & $1.78_{-0.48}^{+0.42}$ & $3.15_{-0.27}^{+0.16}$ & $3.25_{-1.66}^{+5.20}$ & $3.14_{-1.27}^{+3.07}$ & 153.28/151 \\
				0606322101 & $9.48_{-0.70}^{+0.81}$ & $1.32_{-0.09}^{+0.08}$ & $0.79_{-0.02}^{+0.02}$ & & $0.61_{-0.01}^{+0.01}$ & $1.42_{-0.18}^{+0.20}$ & & $0.154_{-0.009}^{+0.009}$ & $1.31_{-0.13}^{+0.14}$ & & $0.888_{-0.235}^{+0.009}$ & $<0.02$ & $2.41_{-0.49}^{+0.59}$ & & $6.406_{-0.027}^{+0.030}$ & $<0.08$ & $1.58_{-0.35}^{+0.36}$ & & $1.89_{-0.37}^{+1.11}$ & $2.70_{-0.19}^{+0.21}$ & $8.77_{-2.38}^{+3.23}$ & $1.13_{-0.54}^{+0.47}$ & 173.59/149 \\
				0606322201 & $6.27_{-0.36}^{+0.37}$ & $1.17_{-0.16}^{+0.10}$ & $0.78_{-0.01}^{+0.01}$ & & $0.74_{-0.01}^{+0.01}$ & $1.06_{-0.27}^{+0.59}$ & & $0.125_{-0.005}^{+0.005}$ & $3.83_{-0.54}^{+0.65}$ & & $0.891_{-0.021}^{+0.015}$ & $<0.08$ & $3.18_{-1.49}^{+3.63}$ & & $6.426_{-0.040}^{+0.039}$ & $0.10_{-0.06}^{+0.05}$ & $2.49_{-0.64}^{+0.64}$ & & $1.97_{-0.49}^{+0.23}$ & $3.45_{-0.22}^{+0.35}$ & $1.70_{-0.54}^{+0.99}$ & $4.45_{-1.40}^{+1.46}$ & 236.97/168 \\
				0606322301 & $16.54_{-0.74}^{+0.81}$ & $1.95_{-0.14}^{+0.06}$ & $0.62_{-0.02}^{+0.02}$ & & $0.74_{-0.01}^{+0.01}$ & $2.13_{-0.43}^{+0.73}$ & & $0.124_{-0.002}^{+0.002}$ & $4.01_{-0.26}^{+0.22}$ & & $0.843_{-0.020}^{+0.015}$ & $0.07_{-0.02}^{+0.02}$ & $9.94_{-2.71}^{+3.52}$ & & $6.442_{-0.031}^{+0.031}$ & $0.14_{-0.03}^{+0.03}$ & $4.12_{-0.66}^{+0.68}$ & & $1.87_{-0.14}^{+0.31}$ & $3.88_{-0.12}^{+0.06}$ & $6.36_{-3.11}^{+16.75}$ & $1.25_{-0.04}^{+4.64}$ & 254.57/169 \\
				\hline
			\end{tabular}
			\begin{tablenotes}
				\footnotesize
				\item In this Table, Column (1): XMM-Newton observational Observational ID; Column (2-4): Parameters of $zxipcf$ model; Column (5-8): Parameters of $compTT$ component; Column (9-11): the first $gaussian$ emission component around $0.9\,\rm keV$; Column (12-14): the second $gaussian$ emission component for Fe {\footnotesize K$\alpha$} line; Column (15-19): Parameters of $relxill$ model; Column (20): Reduced $\chi ^2$ and degree of freedom.
			\end{tablenotes}
		\end{threeparttable}
	\end{table}%
\end{landscape}

\begin{figure*}
	\centering
	\includegraphics[width=0.28\textwidth]{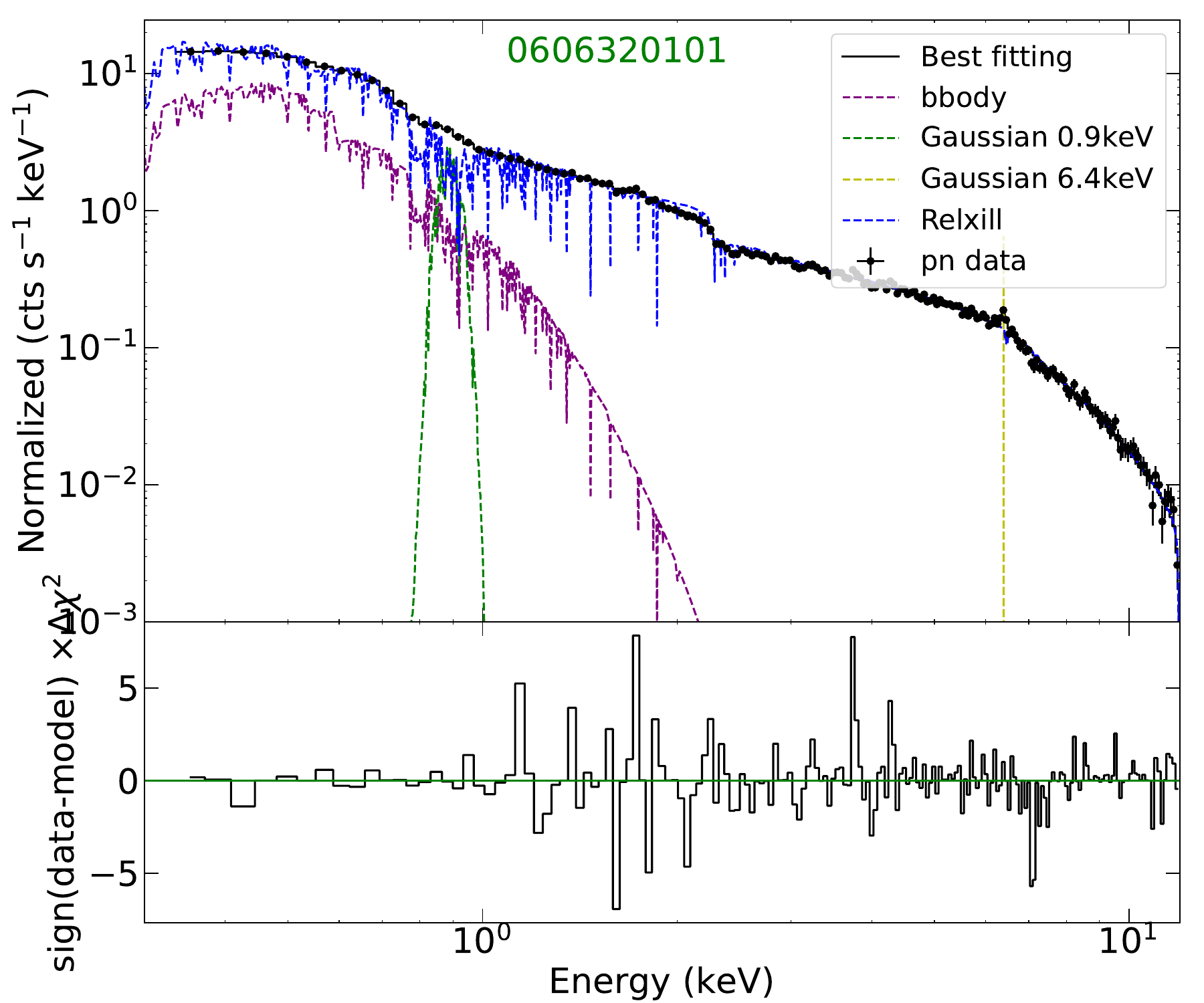}
	\includegraphics[width=0.28\textwidth]{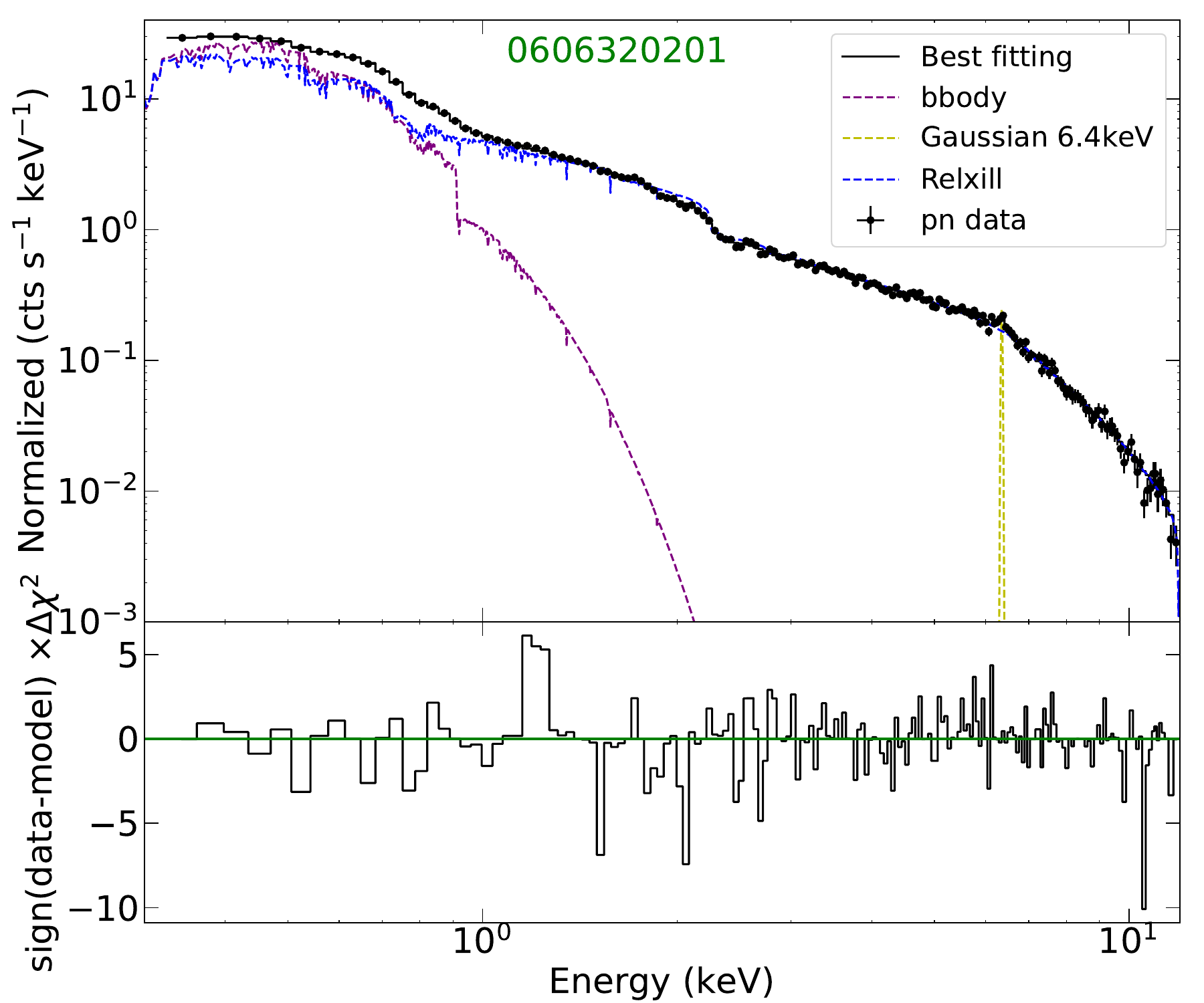}
	\includegraphics[width=0.28\textwidth]{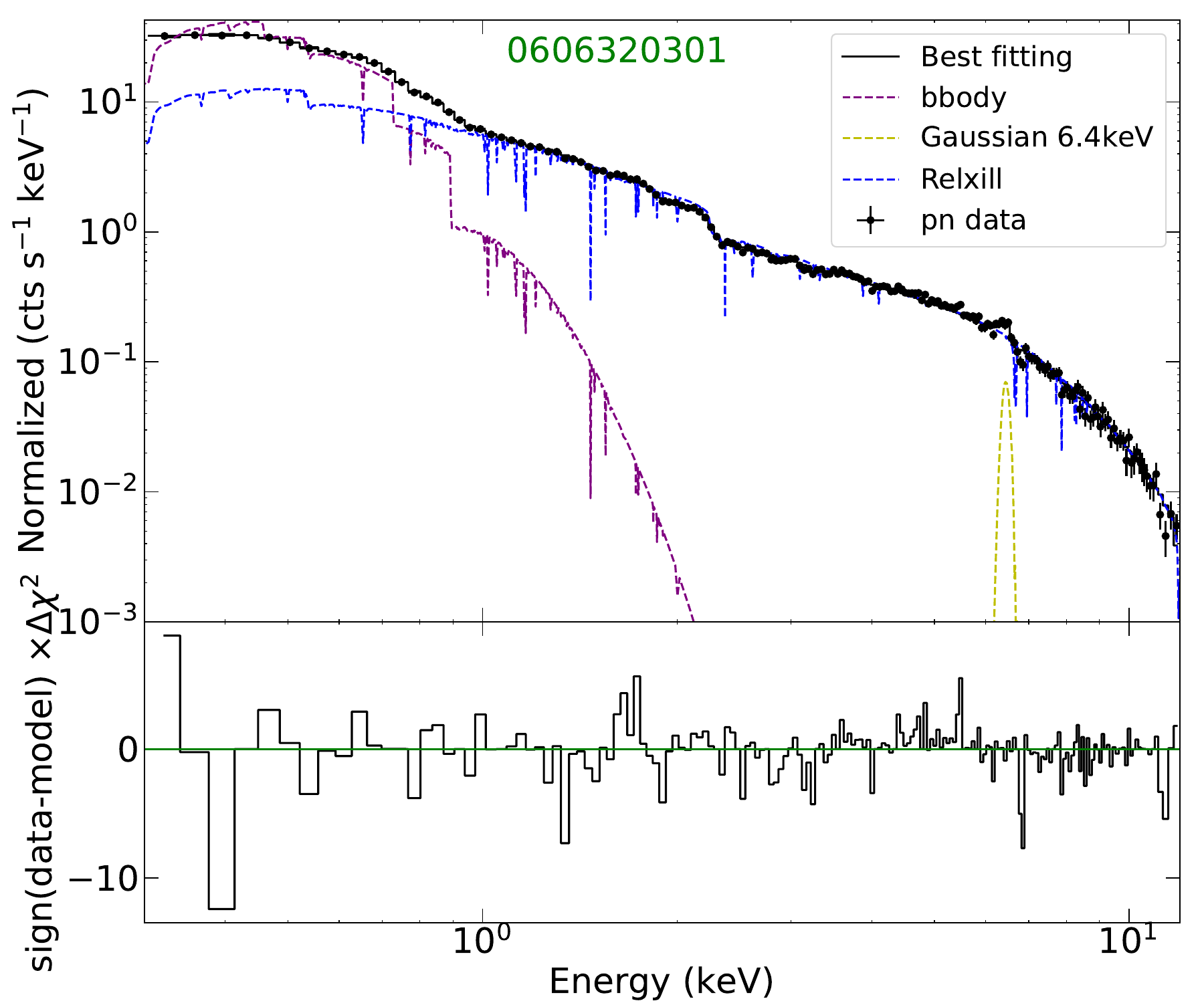}
	\includegraphics[width=0.28\textwidth]{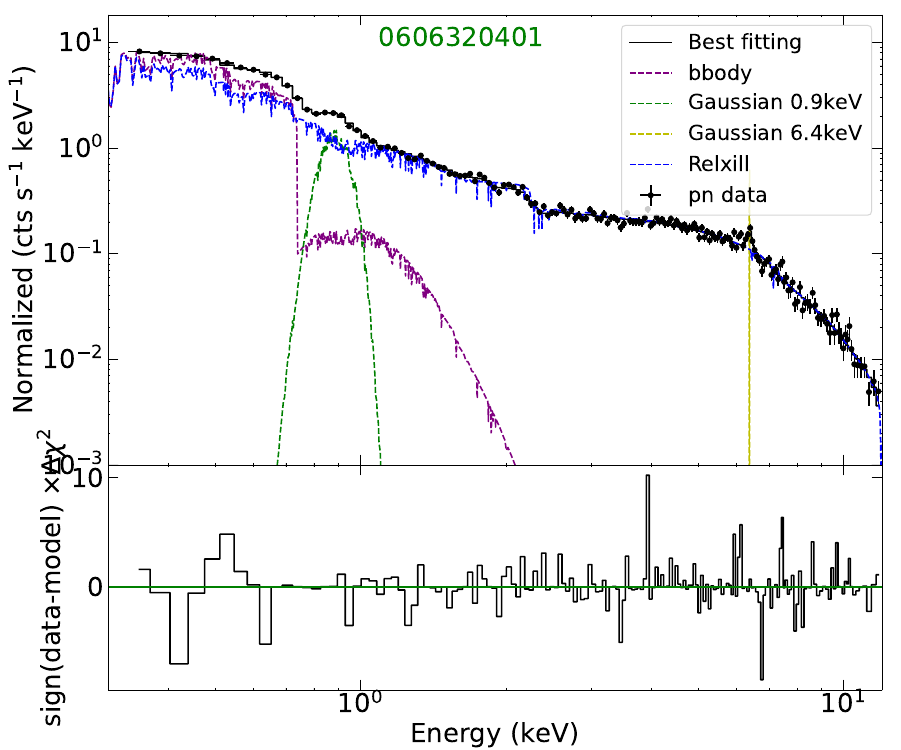}
	\includegraphics[width=0.28\textwidth]{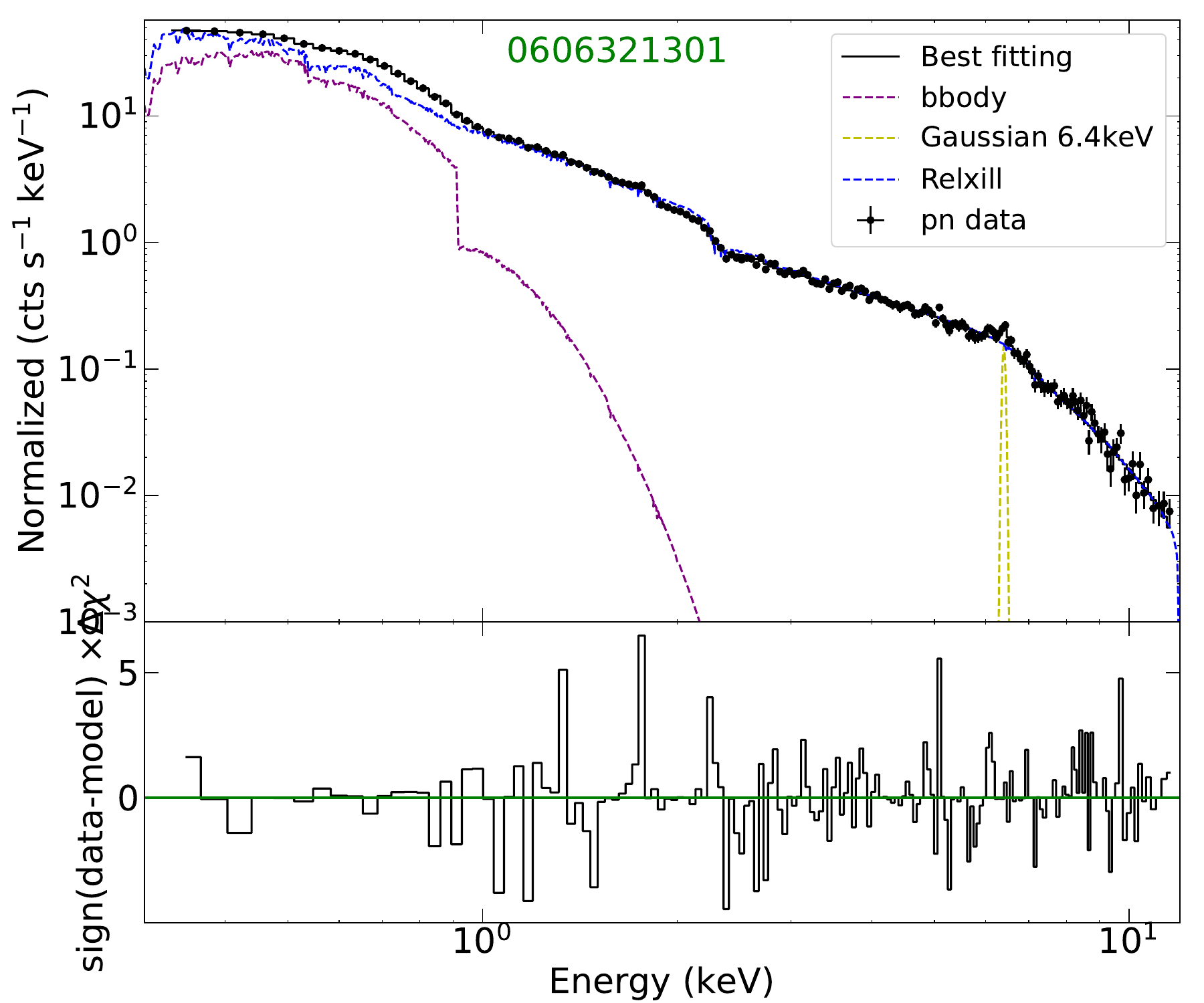}
	\includegraphics[width=0.28\textwidth]{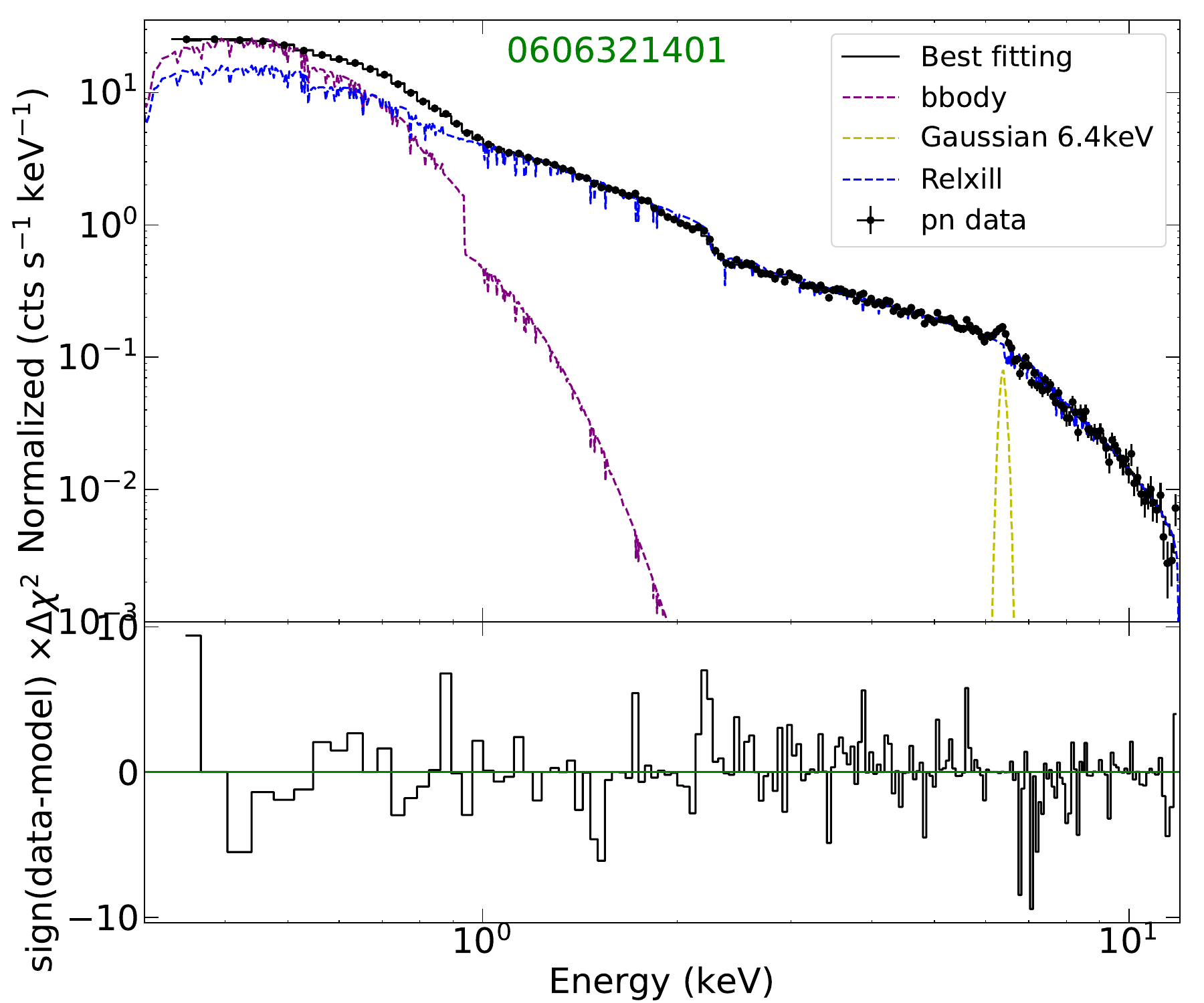}
	\includegraphics[width=0.28\textwidth]{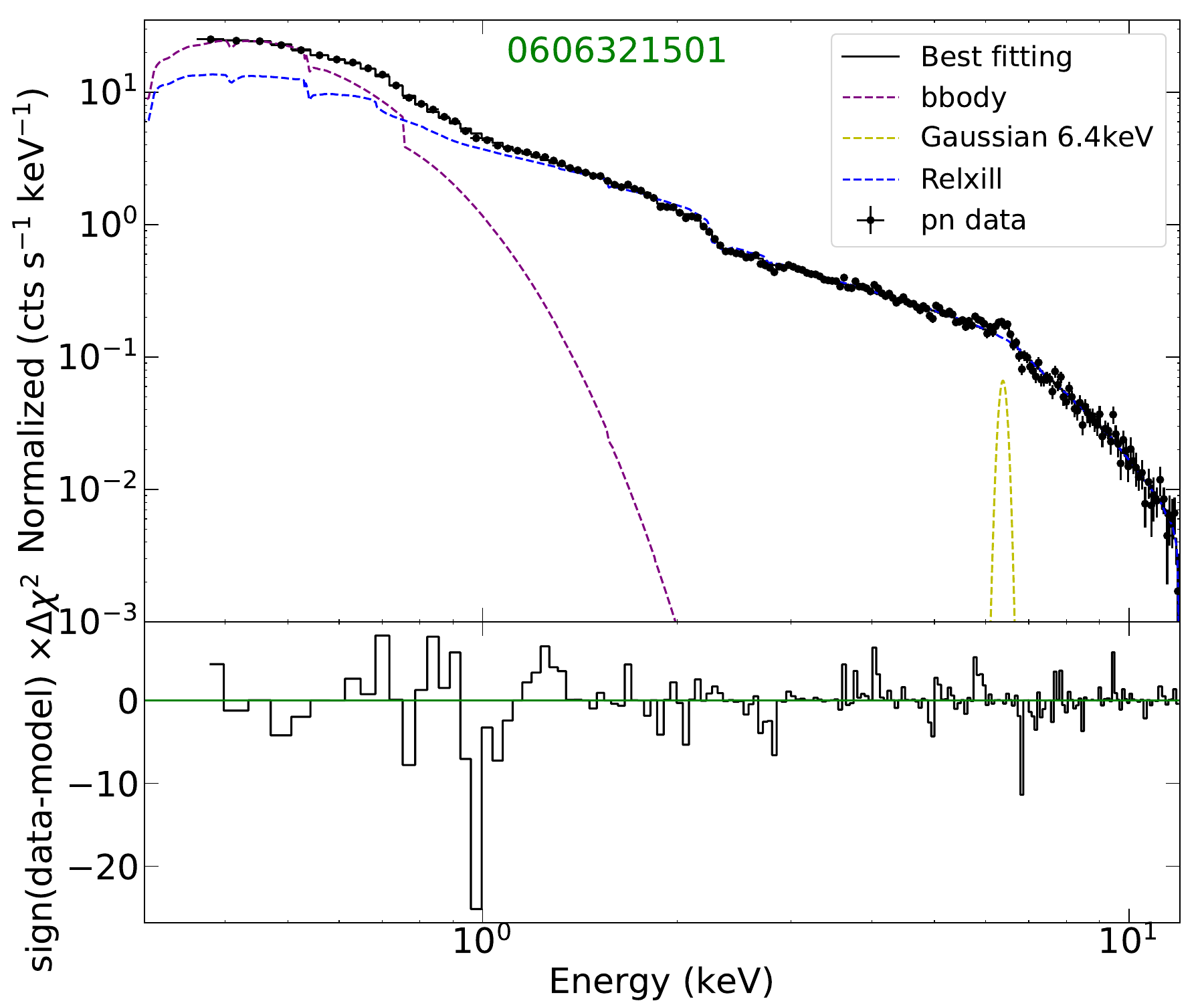}
	\includegraphics[width=0.28\textwidth]{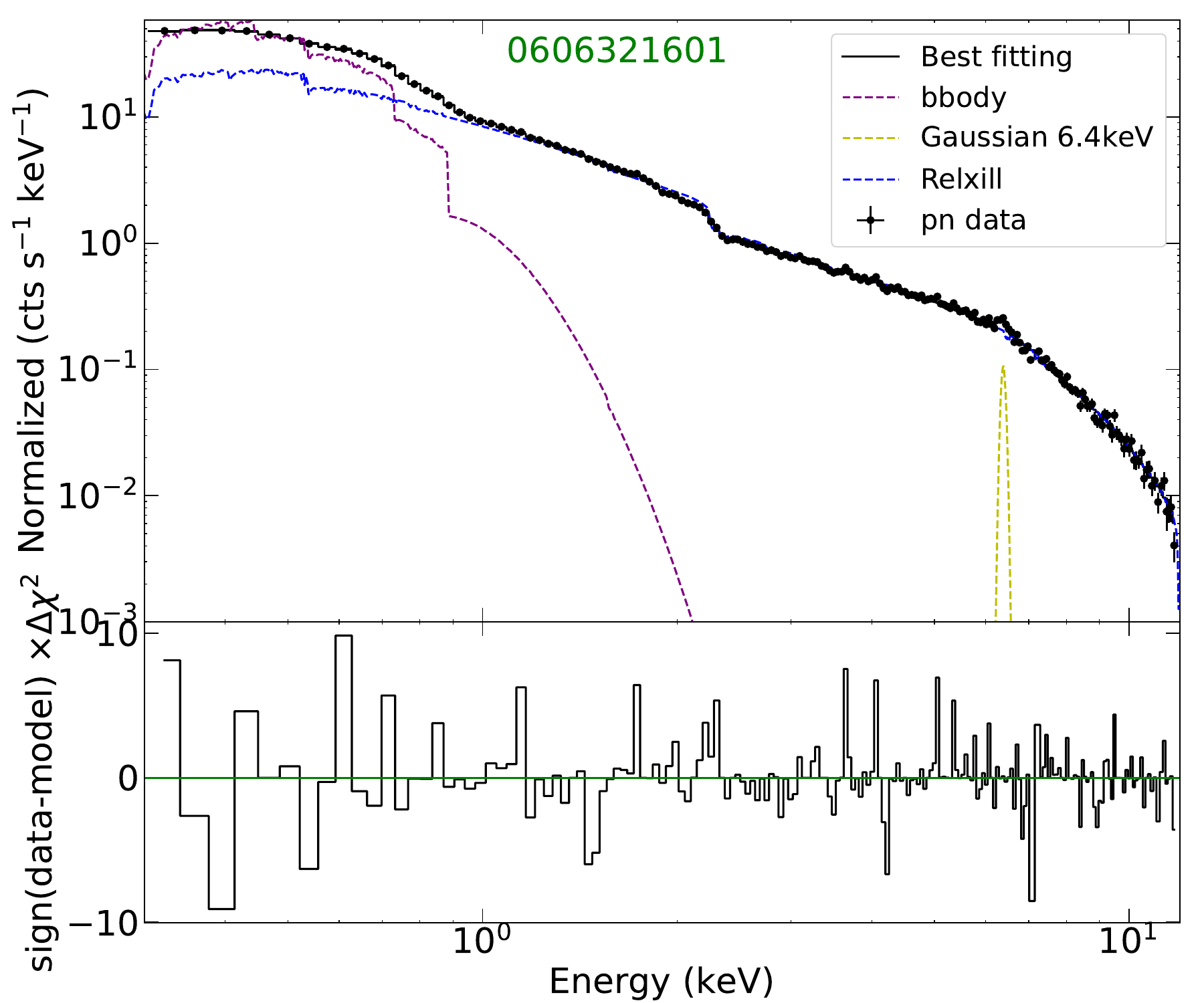}
	\includegraphics[width=0.28\textwidth]{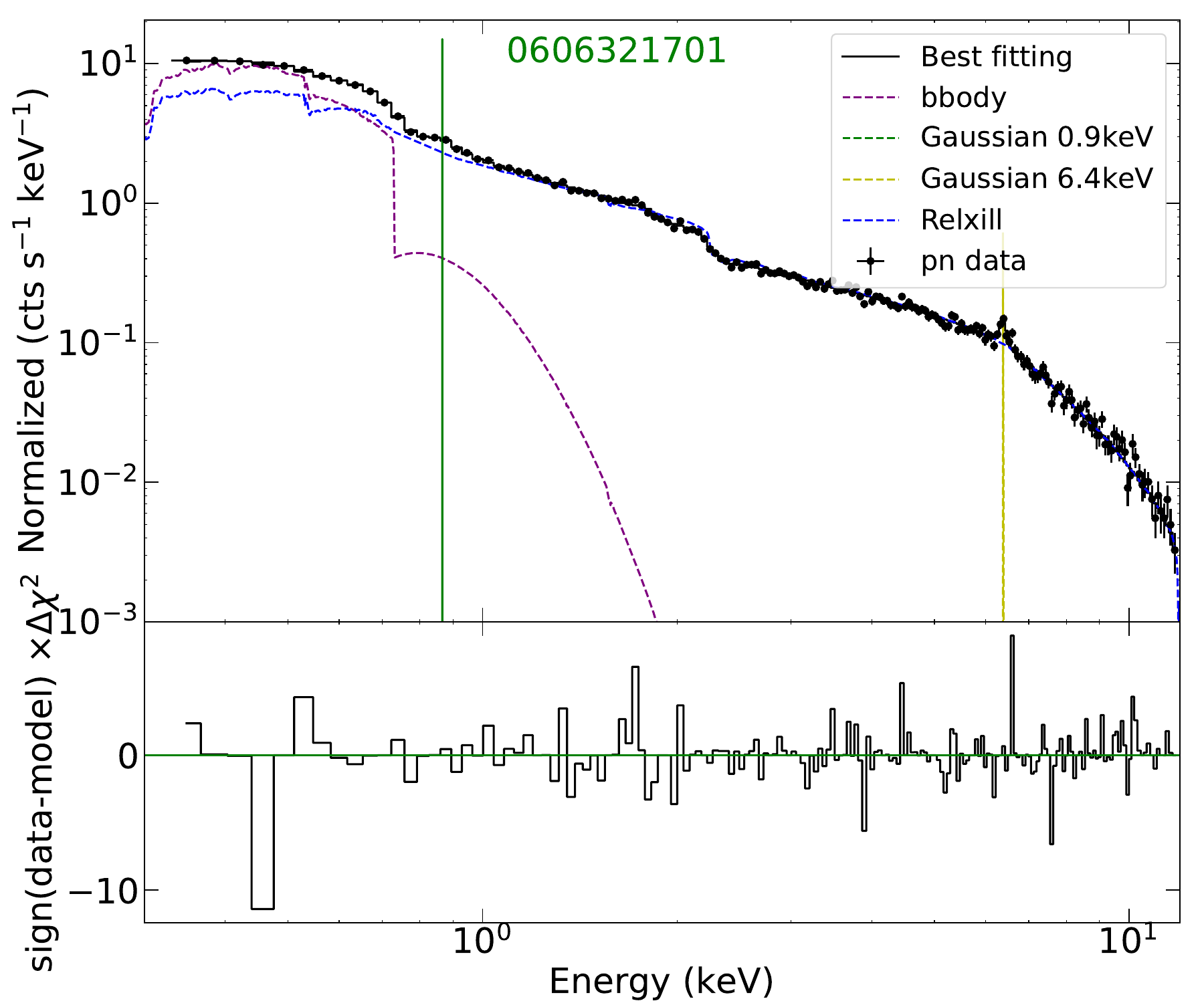}
	\includegraphics[width=0.28\textwidth]{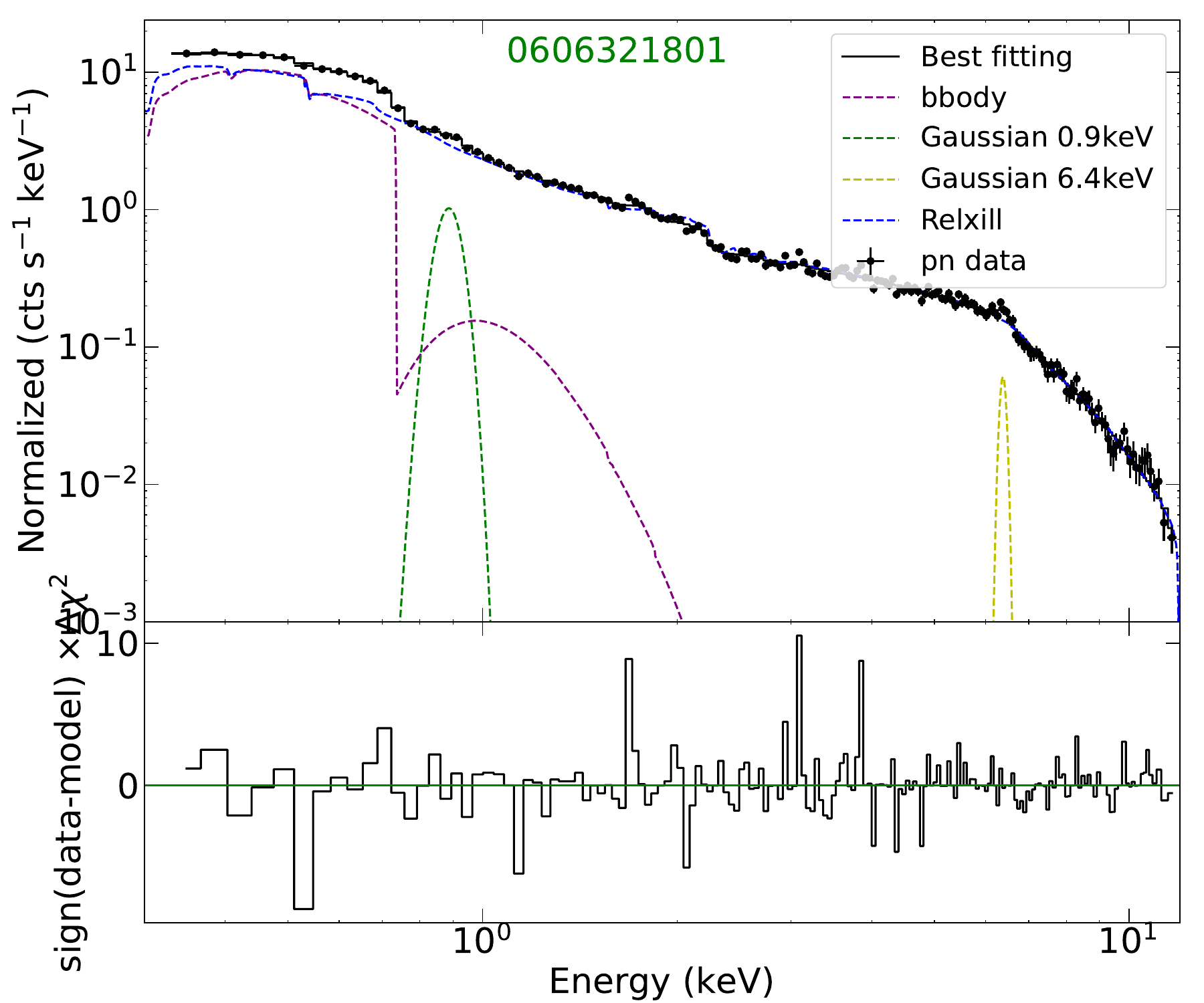}
	\includegraphics[width=0.28\textwidth]{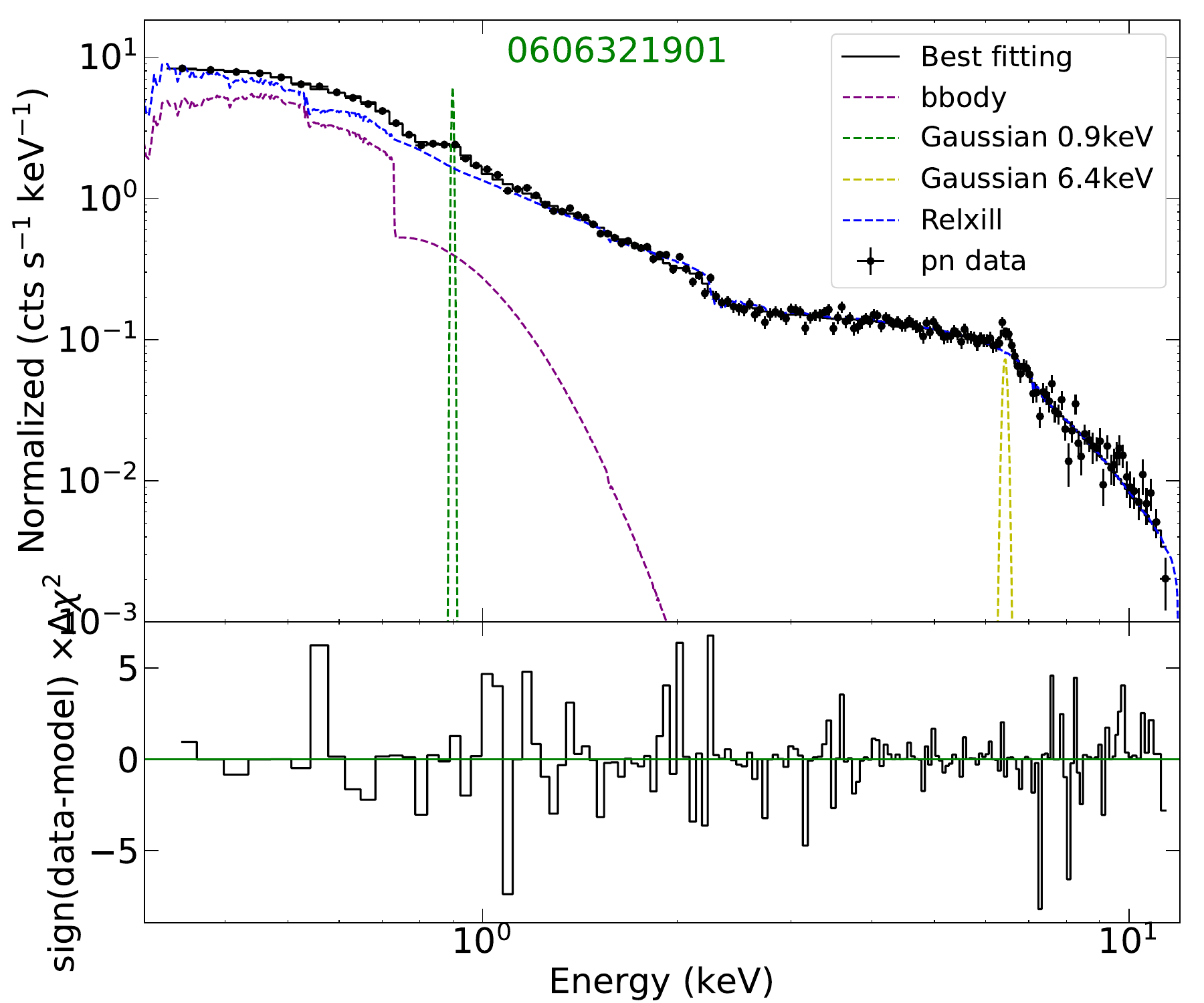}
	\includegraphics[width=0.28\textwidth]{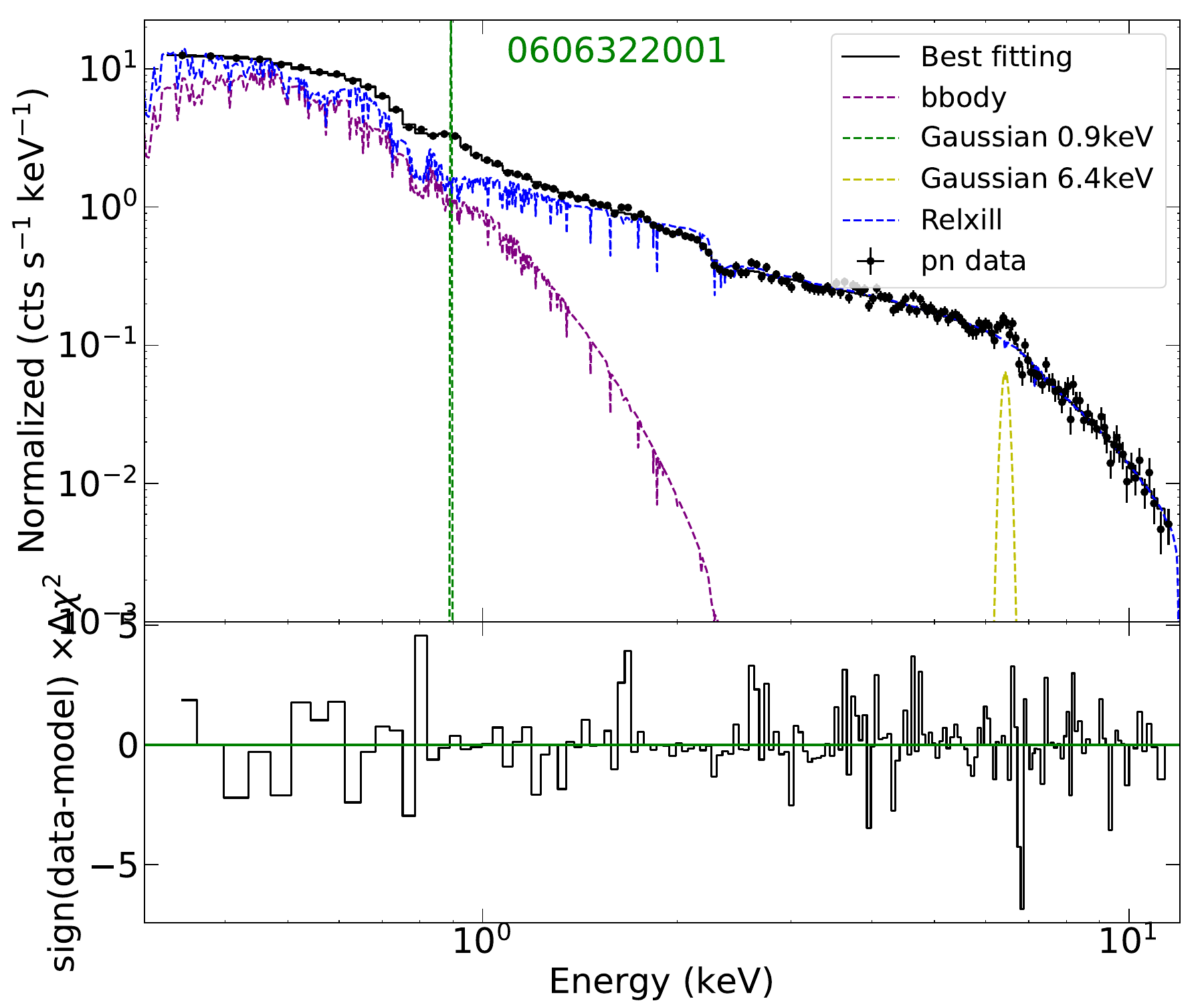}
	\includegraphics[width=0.28\textwidth]{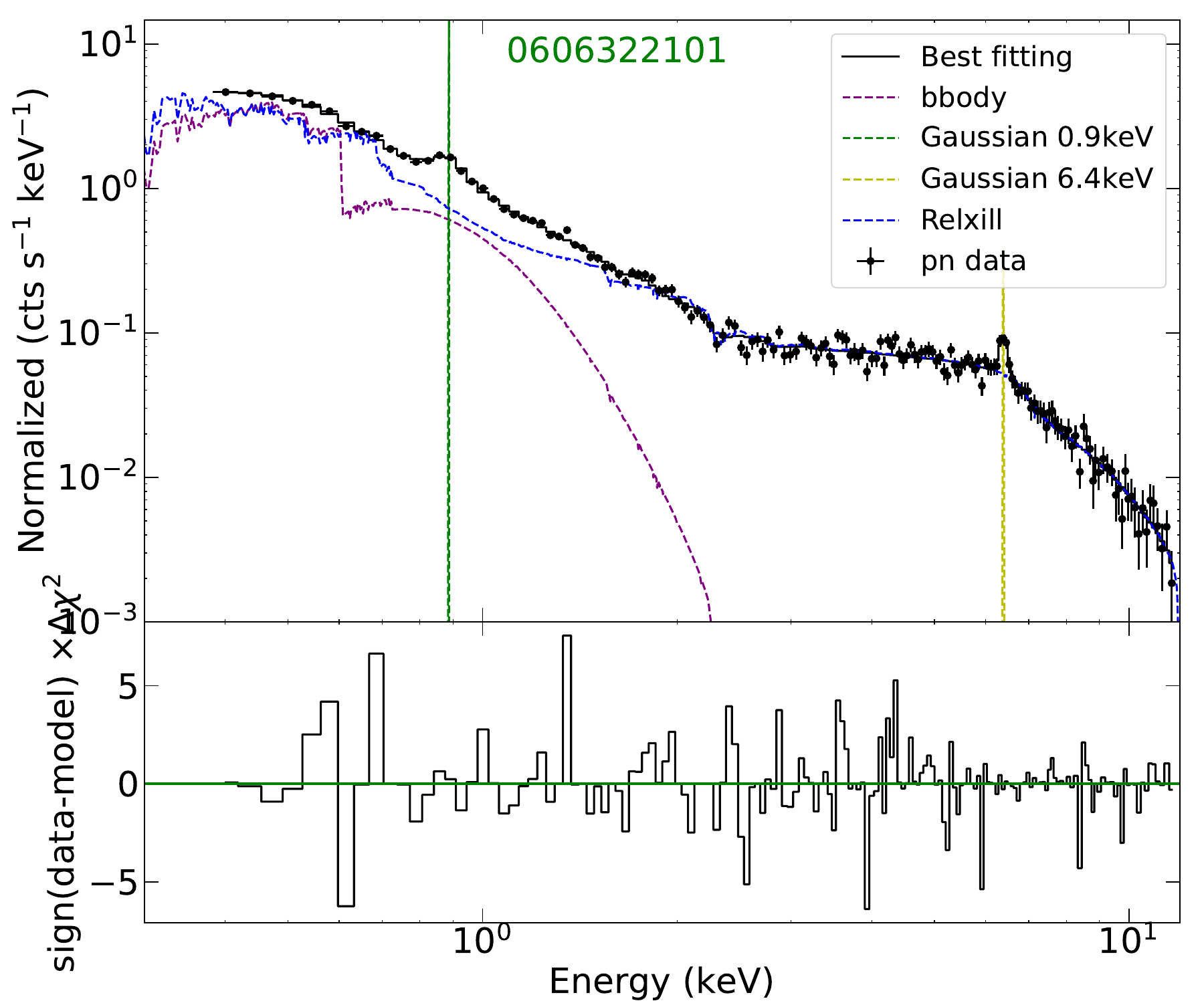}
	\includegraphics[width=0.28\textwidth]{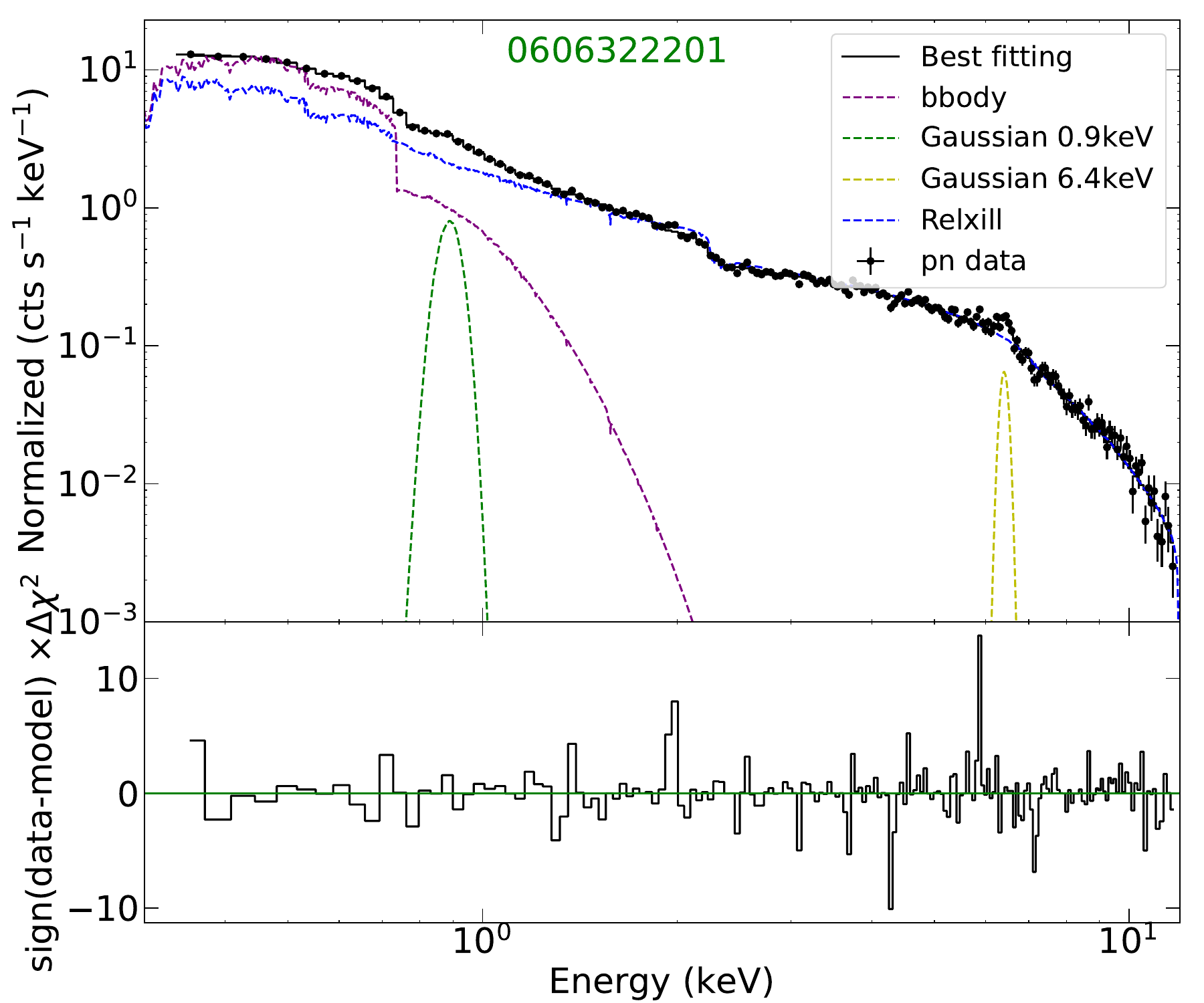}
	\includegraphics[width=0.28\textwidth]{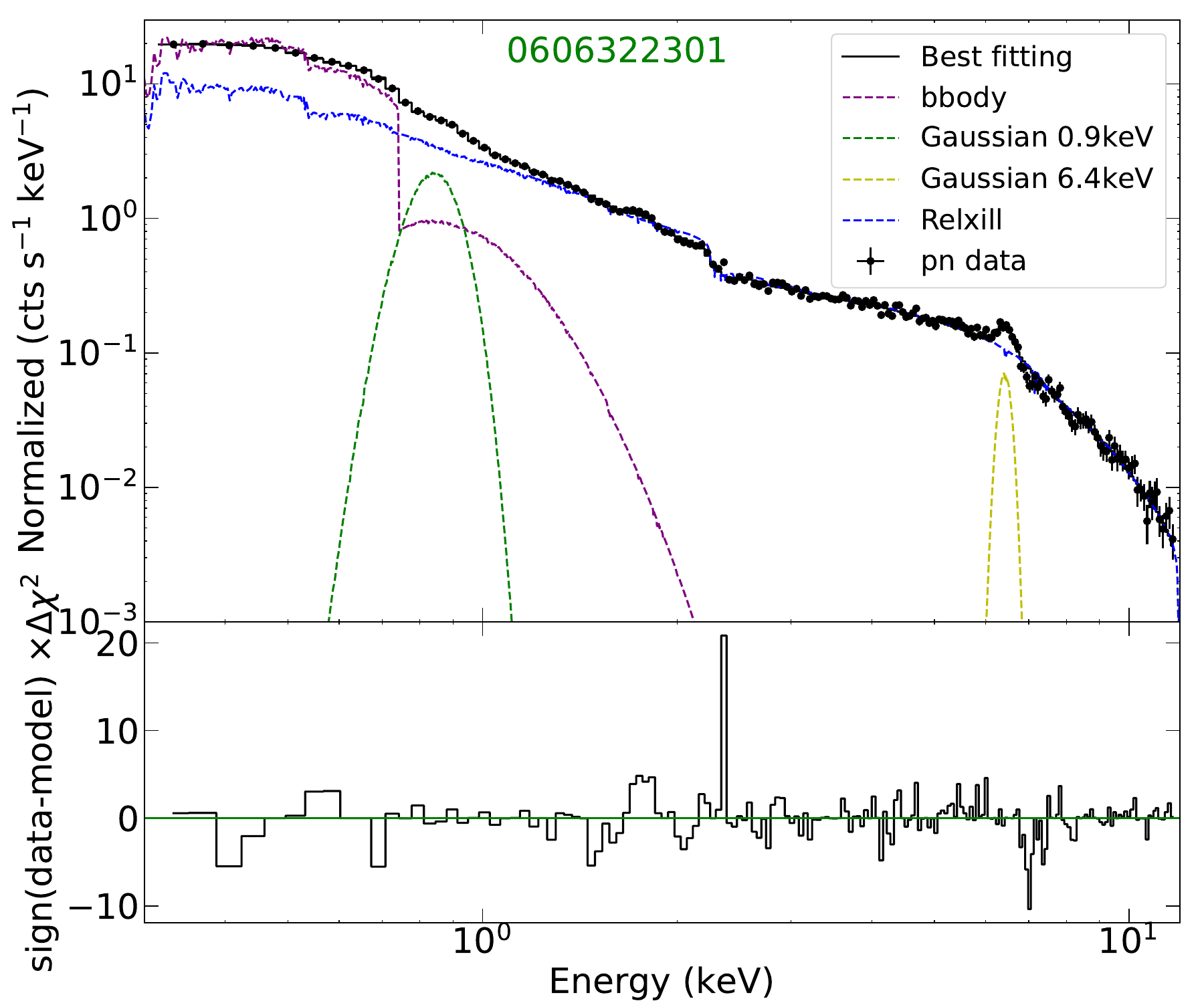}
	\caption{Best model ($phabs*zxipcf*(zedge*bbody+zgauss+relxill+zgauss)$) fitting for 15 XMM-Newton pn spectra of NGC 4051.}
	\label{fig:xmm-spec}
\end{figure*}

When extended to $0.3-2.0~\rm keV$, the residual between the data and the fitted $relxill$ model showed that all spectra have prominent soft X-ray excess components below $2~\rm keV$. 
Considering the $relxill$ could explains the soft-excess \cite[e.g.,][]{2018MNRAS.479.2464G, 2025ApJ...981..186G}, we attempted to fit the soft X-ray excess component with release $R_{\rm in}$, $R_{\rm br}$, and $Index1$. 
However, likely constrained by the relatively narrow energy range ($0.3-12 \rm \, keV$) and affected by the soft X-ray excess, the simple $relxill$ model cannot adequately fit the full-energy-range X-ray spectra. The results show a steeper photon index (generally $>2.5$) in the power-law component and a larger reflection fraction compared to the results for hard X-ray spectra, and greater dispersion of data to model residuals in the soft X-ray energy band. 
Thus, we added a black body model to fit the soft X-ray excess component. 
A portion of the 15 soft X-ray excess spectra can be successfully fitted by the absorbed blackbody model, while the spectral residuals of the other part show an emission excess around $0.9\, \rm keV$, which prompted us to incorporate the second Gaussian component. 
Our final fitting results are shown in Table \ref{tab:xmm_fitting} with the model of $phabs \times zxipcf \times (zedge \times bbody+zgauss+zgauss+relxill)$. 
Figure \ref{fig:xmm-spec} demonstrates the detailed spectral fitting results.

The fitting results of $zxipcf$ parameters in Table \ref{tab:xmm_fitting} reveal consistent hydrogen column density ($nH$) across epochs. However, the ionization parameters ($\log \xi$) and covering fraction ($CF$) after May 27 showed a significant increase, supporting obscuration effects on soft X-ray emission. 
Taking into account the soft X-ray excess components, the fitting results of the three components—$zedge$, $bbody$, and $0.9\,\rm keV$ $zgauss$—show a certain degree of correlation. Several observations in the low X-ray flux state in Figure \ref{fig:swift} upper panel, including seven observations after May 27 and two earlier data (0606320101 and 0606320401), reveal an emission component around $0.9\,\rm keV$, which requires an additional zgauss component. In comparison to the high X-ray flux state, the $bbody$ temperature is slightly higher, while the $zedge$ threshold energy is slightly lower. 
Our results indicate that the high-state X-ray spectrum of NGC 4051 may also contain a $0.9\,\rm keV$ emission component in the soft X-ray band, but it is overwhelmed and not significantly detected, resulting in differences in the spectral fitting components between the high and low states. 
For $relxill$ and $6.4\, \rm keV$ components in Table \ref{tab:xmm_fitting}, we found no significant changes for these parameters before/after May 27. All of our X-ray spectral photon index fitting values are consistent with the general AGN power-law spectrum \cite[e.g., ][]{2007MNRAS.382..194N, 2021RAA....21....4Z}. 

\begin{figure}
	\centering
	\includegraphics[width=1\columnwidth]{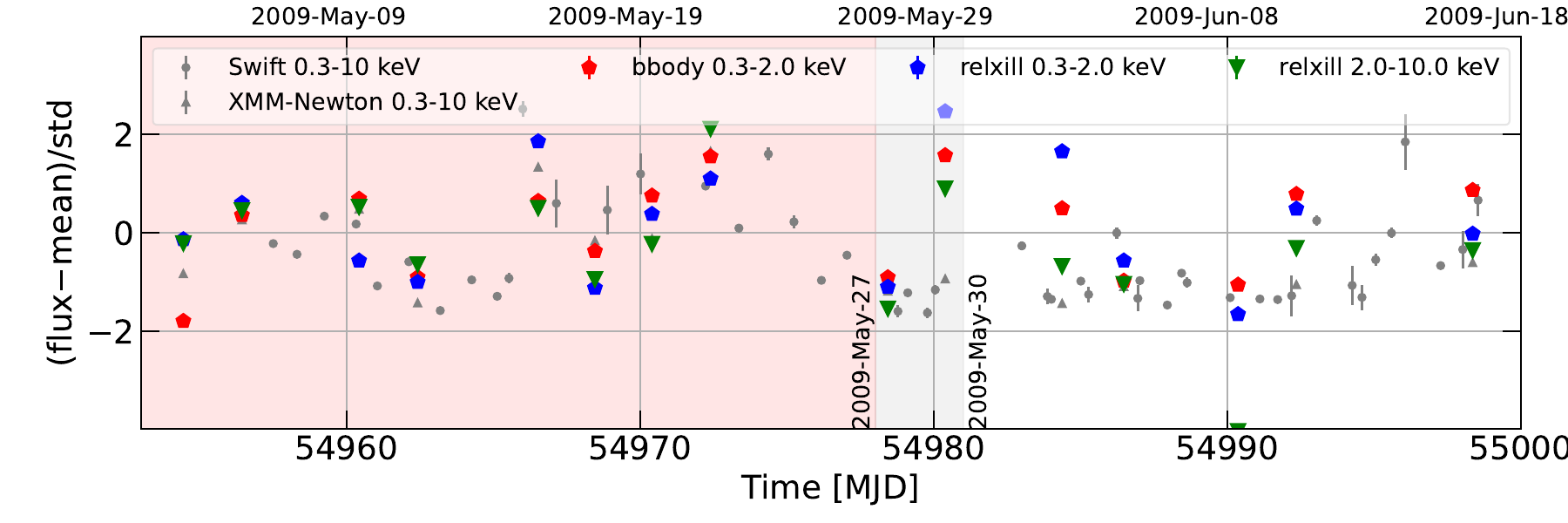}
	\caption{The spectral fitting components flux variation of NGC 4051.}
	\label{fig:xmm_cpts}
\end{figure}

We performed the same analysis on the time-resolved fluxes of spectral components from NGC 4051's XMM-Newton X-ray spectra as in Fig. \ref{fig:swift}, presenting the standard deviation-normalized light curves in Fig. \ref{fig:xmm_cpts}. 
Our analysis shows that both $bbody$ and $relxill$ components exhibited flux variations before May 27 that matched the Swift and XMM-Newton X-ray light curves, validating the reliability of our spectral modeling. 
After May 27, while the $relxill$ hard X-ray light curve remained consistent with Swift's overall trend, the soft X-ray fluxes from both $bbody$ and $relxill$ components generally exceeded Swift's light curve measurements. 
These results indicate that the $relxill$ hard X-ray fluxes entered an intrinsic low state after May 27, whereas the observed suppression in soft X-ray fluxes from both components likely resulted from increased absorption effects.
Only the data of the observation ID 0606321801 shows extremely high soft and hard X-ray flux components relative to the Swift light curve. The high soft X-ray flux deviating from the Swift light curve may be due to the absorption effect, while its high hard X-ray flux contains more reflection component.

\subsection{RGS spectra}
\label{subsec:rgs}

\begin{figure}
	\centering
	\includegraphics[width=1\columnwidth]{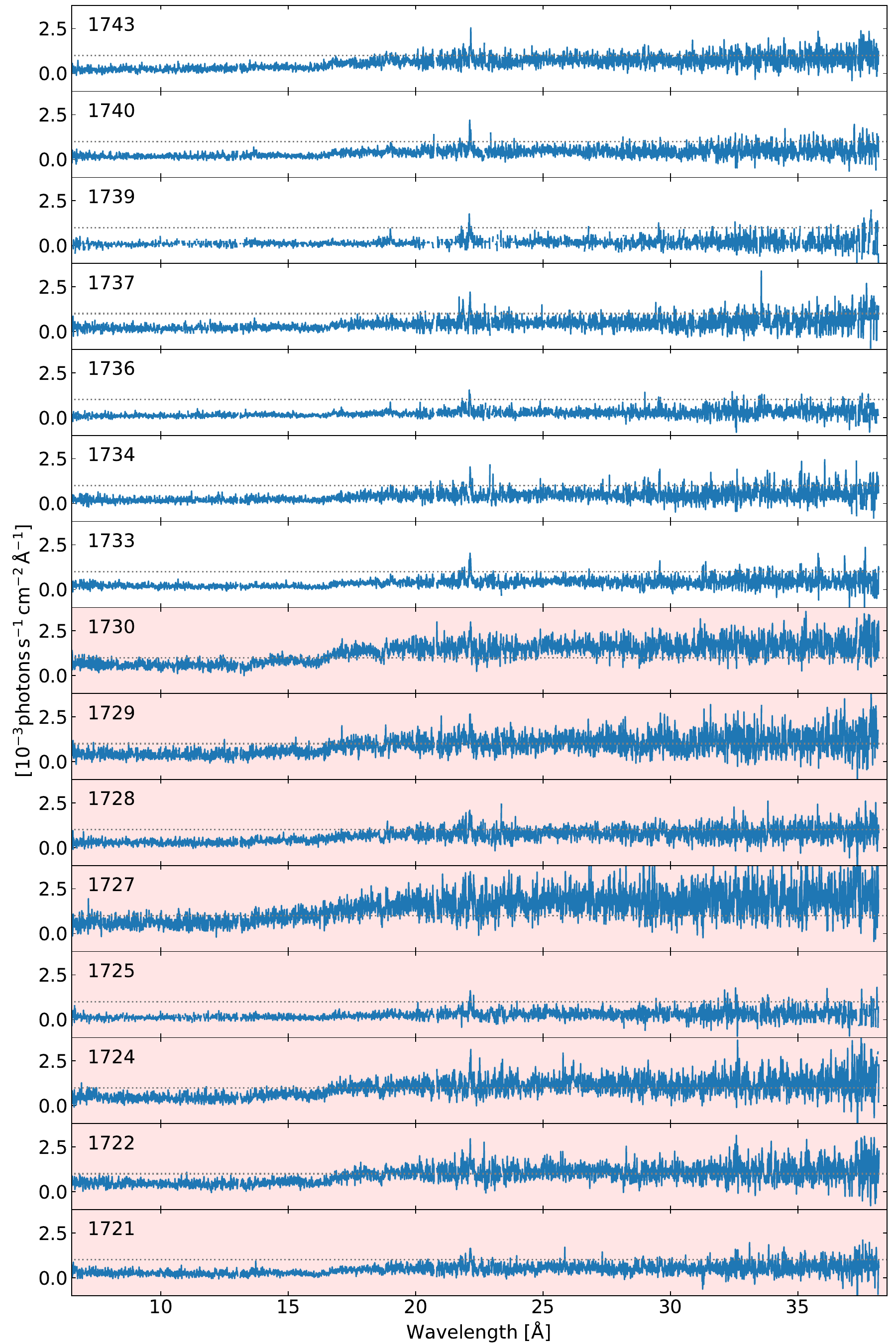}
	\caption{The fluxed RGS spectra of NGC 4051.}
	\label{fig:xmm-rgs}
\end{figure}

Using XMM-Newton RGS data in 2009, \cite{2011MNRAS.413.1251P, 2011MNRAS.415.2379P} discussed the absorption and emission features of NGC 4051, respectively. \cite{2011MNRAS.413.1251P} integrated the RGS spectra at representative high (revs 1727$-$1730) and low (revs 1725, 1736, 1739) flux levels, and found a dominant absorption line spectrum seen against the stronger continuum when bright, and an emission line spectrum emerging more clearly when the continuum is faint. 
Their four high-flux data and one low-flow data (rev 1725) were all observed before May 27, while the remaining two low-flux data were observed subsequent to this date. 
To systematically investigate the soft X-ray spectral characteristics of NGC 4051 during distinct optical-X-ray correlation epochs, we analyze its complete set of 15 RGS observations (Figure \ref{fig:xmm-rgs}), notwithstanding the reduced signal-to-noise ratio in individual exposures when compared with the integrated data product.
In each of the panels in Figure \ref{fig:xmm-rgs}, a reference line with a value of $1\times \rm 10^{-3} photons\, s^{-1}\, cm^{-2}$ \AA$^{-1}$ is marked by a dotted line to visually compare the flux changes. 
Consistent with the light curve results shown in Figure \ref{fig:swift}, the flux values of almost all RGS spectra in the long-wave band ($\lambda > 16.8$ \AA, with the exception of the observations with rev of 1721 and 1725) before May 27 were above the reference line of $1\times \rm 10^{-3} photons\, s^{-1}\, cm^{-2}$ \AA$^{-1}$ that all spectra observed after that date were generally paused. 

Consistent with integrated RGS spectra results reported by \cite{2011MNRAS.413.1251P}, all analyzed RGS spectra exhibit a prominent absorption edge near 16.8 \AA\ accompanied by an O {\footnotesize VII} forbidden line at 22.101 \AA. The absorption edge around 16.8 \AA\ could be the result of ionized absorption \cite[e.g.,][]{2008MNRAS.386L...1S}.
While only the high-flux RGS spectra have an absorption line at 18.67 \AA\  which could be the absorption component of O {\footnotesize VIII} Lyman $\alpha$ (18.97 \AA) with a line-of-sight outflow velocity of about $-4744\, \rm km/s$. 
In addition to the pre-May 27 RGS spectrum, the observation from May 27 (rev 1733) also reveals a weak absorption line at 18.67 \AA. 

\section{Discussion}
\label{sec:discu}

\subsection{Main results}
\label{subsec:MainResults}

By analyzing quasi-simultaneous Swift and XMM-Newton optical/X-ray observations, we identified distinct differences in the optical-X-ray flux correlation of NGC 4051 around May 27, 2009. Before this date, optical and X-ray fluxes showed coordinated variations, whereas after May 27, optical flux was significantly enhanced while X-ray photon count rate remained relatively low. 
Furthermore, short-timescale X-ray variability and RGS spectra displayed differences before/after May 27. Pre-May 27 observations showed larger soft/hard X-ray amplitude variations than subsequent observations, which were relatively stable. And, more significant absorption features were detected in RGS spectra before May 27. 
With spectral fitting, the observations in post-epochs have a larger absorption covering fraction.  

These results suggest an abrupt change in NGC 4051's optical/X-ray emission mechanisms during May-June 2009, driving shifts in multi-band correlations and altered short-timescale variability properties.

From an alternative analytical perspective, while discernible spectral variations emerged across the May 27 demarcation, subsequent X-ray spectral analysis demonstrated no statistically significant deviations from prior low-state observations (i.e., rev 1721 and 1725). 
Cross-examination of post-May 27 datasets (seven observations) with these two baseline epochs consistently demonstrated two characteristic features: a persistent $0.9\ \rm keV$ emission component (Figure \ref{fig:xmm-spec}) and enhanced absorption covering fractions.
Additionally, XMM-Newton soft X-ray variability amplitudes in revs 1721 and 1725 were notably lower than those in pre-May 27 observations (Table \ref{tab:xmm}). 

\subsection{Possible explanations}
\label{subsec:PossibleExp}

NGC 4051 exhibits strong long-term time-averaged optical-X-ray flux correlations \cite[][]{2000MNRAS.312..880U, 2003MNRAS.343.1341S, 2010MNRAS.403..605B, 2024MNRAS.527.5668K}, yet shows no evidence of rapid optical variability analogous to its X-ray fluctuations \cite[][]{2000ApJ...542..161P}. 
\cite{2003MNRAS.343.1341S} proposed that the complex correlation might originate from a combination of X-ray reprocessing and propagating disturbances in the outer-to-inner accretion disk regions.
Based on the optical and X-ray light curves spanning over 12 years, \cite{2010MNRAS.403..605B} pointed out that the X-ray reprocessing model cannot account for all the optical variability, and the $38.9^{2.7}_{8.4} \, \rm d$ optical delay might be attributed to the reflection process in the dust. 
NGC 4051's good optical-X-ray correlation before May 27 consistent with historical literature, while the following optical flares with no significant X-ray variation challenges reprocessing process and thermal Comptonization model, requiring further investigation.

Spectral analysis reveals a pronounced $0.9\, \rm keV$ excess emission component in low X-ray flux states of NGC 4051, modeled as a Gaussian feature. \cite{2008PASJ...60.1257H} attributed this component to circumnuclear starburst activity.
Starburst galaxies exhibit thermal luminosities comparable to AGN \cite[][]{1998ApJ...505..174G, 2007ApJ...671.1388D}. If the post-May 30 UV brightening in NGC 4051 originates the starburst, which also predicts enhanced soft X-ray emission \cite[][]{2001ApJ...550..230L}. However, this contradicts our observed soft X-ray light curves. Furthermore, the post-May 27 X-ray spectra (sharing the $0.9\ \rm keV$ excess) show no significant spectral model differences as the pre-May 27 low-state observations rev 1721 and 1725.

\cite{2023MNRAS.519..909Z} discussed the presence of jet might explain optical-X-ray uncorrelations in a radio-loud quasar SDSS J121426.52+140258.9. 
Many sample studies have suggested that the jet may also play an important role in the X-ray production for non-blazar AGNs \cite[e.g.,][]{2011ApJ...726...20M, 2017ApJ...835..226K}. 
\cite{2011MNRAS.412.2641J} studied quasi-simultaneous radio (VLA)-X-ray (RXTE) observations of NGC 4051 over 16 months starting in 2000-01. 
Though NGC 4051 exhibits three discrete radio components at 8.4 GHz in VLA-A configuration, no visible radio-X-ray correlations were detected.
\cite{2011ApJ...729...19K} analyzed quasi-synchronous radio (VLA)-X-ray (Chandra) data for NGC 4051, identifying a compact jet in 8.4 GHz observations. After removing extended radio emission effects, they found an X-ray-anticorrelated radio component consistent with disk-jet coupling.
\cite{2017MNRAS.465.1336J} reanalyzed 2008 data accounting for VLA beam configuration effects, reaffirming no significant nuclear radio variability. They proposed that NGC 4051's jet structure might represent earlier activity remnants, with current nuclear jet emission being extremely weak.
This interpretation aligns with parsec-scale Very Long Baseline Array (VLBA) non-detections in 2005 \cite[][]{2013ApJ...765...69D}. Conversely, 2007 EVN observations revealed three 1.6 GHz sub-arcsecond components and a nuclear-dominated 1.5 GHz source with a steep spectral index (\( 0.7\pm0.1 \)) and brightness temperature \( T=2\times10^5\,\rm K \), favoring outflow scenarios over relativistic jets \cite[][]{2009ApJ...706L.260G}. 
Given NGC 4051's confirmed large-scale jet structure, its post-May 27, 2009 UV brightening and X-ray suppression might originate from jet activity: enhanced nuclear jet emission may suppress coronal X-rays while boosting UV radiation via jet-related processes.
We attempted to process the 2011 VLBA $22\,\rm GHz$ observation data of NGC 4051, but no significant radio signal was detected. The influence of jet on optical and X-ray emission cannot be directly demonstrated.

The absence of correlations between extreme ultraviolet (EUV) and optical/UV bands in NGC 5548 \cite[][]{2016ApJ...824...11G} may be explained by the coronal collapse model \cite[where the accretion disk obscuration rate increases, ][]{2018ApJ...857...86S} or the disk wind obscuration model \cite[e.g.,][]{2019ApJ...877..119D}. 
Meanwhile, \cite{2020ApJ...892...63C} proposed that coronal heating is related to turbulence in the inner accretion disk, with results indicating that the X-ray corona resides in the inner EUV region of the accretion disk. 
These obscuring structures may influence the soft X-ray emission and even the X-ray/EUV to optical/UV correlations. 
Although absorption may affect NGC 4051's soft X-ray emission after May 30, a simple obscuration model cannot fully explain the reduction in hard X-ray emission and the associated optical flare. 

The first version of the inhomogeneous disc model \cite[][]{2011ApJ...727L..24D, 2016ApJ...826....7C} described the quasar variability in optical/UV originating from temperature fluctuations in individual zones of the accretion disc. 
To explain the multi-band light curve correlations, \cite{2018ApJ...855..117C} proposed a modified inhomogeneous disk model, which includes a common large-scale fluctuation based on independent local fluctuations, to simulate the effect of turbulent propagation in all directions within the disk. 
When the stochastic nature of turbulent effects dominates in the accretion disk turbulence model of \cite{2018ApJ...855..117C}, particularly when the truncation radius of the accretion disk approaches the typical UV-emitting region, the correlation between EUV and UV emission may weaken significantly. 
\cite{2020ApJ...892...63C} proposed that coronal heating is related to turbulence in the inner accretion disk. Without the light echoing, the inhomogeneous turbulence model can well reproduce the unusual observed UV to X-ray lags in four local Seyfert galaxies \cite[][]{2020ApJ...892...63C} and the observed stochastic relation between X-ray and UV variations \cite[][]{2022MNRAS.512.5511S}. 

Based on the above discussion, we find that the inhomogeneous turbulence model is more capable of explaining the optical and X-ray correlations of NGC 4051 during May-June 2009 compared to other models, such as the starburst process, jet process, or single obscuration model. 
The inhomogeneous turbulence model suggests that the time lags in the optical/UV and X-ray bands of AGN are due to the perturbation process of the accretion disk, which can explain the good UV-X-ray correlation of NGC 4051 before May 27. 
If the accretion process of NGC 4051 had a significant change in the accretion rate (UV flare) after May 27, it could cause changes in the corona environment in the inner region of the accretion disk, and subsequently intensify the absorption effect of soft X-ray components. 
The UV and X-ray variations of NGC 4051 after May 27 are similar to the insignificant X-ray changes during the optical flare process of a Tidal disruption event (TDE) \cite[X-ray flares begin few hundreds days later,][]{2017ApJ...851L..47G, 2022MNRAS.511.5328G}, but the timescale of the UV flare does not match that of a normal TDE. It may be analogous to the TDE process, where NGC 4051 accreted a clump of neutral material (like star debris) around May 29, leading to a rapid increase in the accretion rate, a decrease in the ionization degree of X-ray absorption, and an increase in the obscuration rate (outflow). 
Similar to the interaction between TDE and the AGN corona \cite[e.g.,][]{2020ApJ...898L...1R}, the accretion of a large amount of neutral, relatively cold material may reduce the temperature of the corona in NGC 4051, leading to a decrease in the hard X-ray flux from the corona. 
The abnormally high state of each X-ray-fitting component observed on May 29, along with variations in RGS spectral outflow absorption, supports a change in the accretion environment. 
It is precisely because the variations in all bands of NGC 4051 throughout the entire period are due to the same inhomogeneous disk perturbation process that the low-state X-ray emission properties (such as X-ray fitting results, RGS spectra) after May 30 show no significant difference from the two previous low-state observations (revs 1721, 1725). 

\section{Summary}
\label{sec:summary}

This paper conducts a detailed analysis of the quasi-simultaneous optical/UV and X-ray observations of NGC 4051 by Swift and XMM-Newton during May-June 2009. The light curves show that the optical/UV and X-ray emissions were well correlated in the first half of the observation period (before May 27), while in the second half, the optical/UV fluxes significantly increased, while the X-rays remained in a relatively stable low state. Compared with the X-ray emissions in the first period, the second period shows several significant differences: 1) the amplitude of short-term soft and hard X-ray variations decreased; 2) the absorption effect of soft X-rays was more pronounced; 3) the intrinsic flux of the hard X-ray component was lower; 4) the RGS energy spectrum absorption lines were not significant, but there were more obvious emission line features. Compared with the X-ray properties in the second period and the two low-state data in the first period, the above differences were not significant. 
Additionally, the hard X-ray hardness ratio to luminosity relation and the fitted photon index did not show significant differences between the two periods. 

The above results do not conform to the classical X-ray reprocessing model. After excluding the starburst process and simple absorption effects, although the possible jet effect cannot be completely ruled out, our results support the accretion disk perturbation model more. That is, the accretion rate of NGC 4051 increased after May 27, causing an optical/ultraviolet outburst. Due to the sudden change in the accretion process, the normal heating of the inner corona was affected, resulting in more significant soft X-ray absorption, suppression of hard X-rays to a certain extent, etc.

\section*{Acknowledgements}

We thank the anonymous referee and editor for valuable and insightful suggestions that improved the manuscript. 
This work is supported by the Natural Science Foundation of Jiangxi (No. 20232BAB211024), the Doctoral Scientific Research Foundation of Shangrao Normal University (Grant No. K6000449), and the project of Jiangxi Province Key Laboratory of Applied Optical Technology (No. 2024SSY03051). 

\section*{Data Availability}

The data underlying this article are available in NRAO (https://data.nrao.edu/portal), XMM-Newton (https://www.cosmos.esa.int/web/xmm-newton), and Swift (https://www.swift.ac.uk/index.php) data centers.  




\bibliographystyle{mnras}
\bibliography{ms} 




%
%


\bsp	
\label{lastpage}
\end{document}